\documentclass[journal]{IEEEtran}
\usepackage{cite}
\usepackage{amsmath,amssymb,amsfonts}

\usepackage{algorithmic}
\usepackage[ruled]{algorithm2e}
\usepackage{multirow,tabularx}
\usepackage[ruled]{algorithm2e} 
\usepackage{graphicx}
\usepackage{subfig}
\usepackage{caption}
\usepackage{setspace}
\usepackage{epsfig,amssymb}
\usepackage{epstopdf}
\usepackage{amsmath}
\usepackage{times}
\usepackage{mathrsfs}
\usepackage{array}

\newtheorem{theorem}{Theorem}
\newtheorem{remark}[theorem]{Remark}
\usepackage{amsmath, latexsym, amsfonts, amssymb}
\usepackage[mathscr]{eucal}
\usepackage{array, tabularx}
\usepackage[table]{xcolor}
\usepackage{multirow}
\usepackage{epsfig}
\usepackage{collcell}
\usepackage{hhline}

\usepackage{url}
\usepackage{mathtools}
\usepackage{psfragx}
\usepackage{xcolor}
\usepackage{bbm}
\usepackage{booktabs}       
\usepackage{mwe}
\usepackage{graphbox} 
\usepackage{adjustbox}
\usepackage{soul}
\usepackage{textcomp}

\begin{document}
\title{Fixed-Symbol Aided Random Access Scheme for Machine-to-Machine Communications}

\author{
\IEEEauthorblockN{Zhaoji~Zhang, Ying~Li, \emph{Member, IEEE}, Lei Liu, and Wei Hou}
}

\maketitle
\begin{abstract}
The massiveness of devices in crowded Machine-to-Machine (M2M) communications brings new challenges to existing random-access (RA) schemes, such as heavy signaling overhead and severe access collisions. In order to reduce the signaling overhead, we propose a fixed-symbol aided RA scheme where active devices access the network in a grant-free method, i.e., data packets are directly transmitted in randomly chosen slots. To further address the access collision which impedes the activity detection, one fixed symbol is inserted into each transmitted data packet in the proposed scheme. An iterative message passing based activity detection (MP-AD) algorithm is performed upon the received signal of this fixed symbol to detect the device activity in each slot. In addition, the deep neural network-aided MP-AD (DNN-MP-AD) algorithm is further designed to alleviate the correlation problem of the iterative message passing process. In the DNN-MP-AD algorithm, the iterative message passing process is transferred from a factor graph to a DNN. Weights are imposed on the messages in the DNN and further trained to improve the accuracy of the device activity detection. Finally, numerical simulations are provided for the throughput of the proposed RA scheme, the accuracy of the proposed MP-AD algorithm, as well as the improvement brought by the DNN-MP-AD algorithm.
\end{abstract}


%

\section{Introduction}\label{sec:introduction}
\IEEEPARstart{T}{he} emerging Internet of Things (IoT) achieves information exchange for objects in the physical world and enlightens the future development of many areas, such as Industry 4.0 and smart cities, etc. Providing efficient support for the IoT and related services is one of the major objectives for Machine-to-Machine (M2M) communications. In the upcoming IoT era, it is anticipated that the number of M2M devices would exceed 50 billion by 2020 \cite{2020}.  Furthermore, many M2M applications, such as smart metering, e-health, intelligent transportation and fleet management are generally characterized by small-sized data packets intermittently transmitted by a massive number of M2M devices. Specifically, these M2M devices would be activated with a low probability \cite{servicetype,servicetypetwo}, which implies that the random-access (RA) process in M2M communications exhibits prominent features of massiveness and sparseness.
\subsection{Challenges}
Due to the massiveness of the M2M communications, attempts to access cellular networks from a huge number of M2M devices can lead to severe congestion and collision in existing RA schemes. Confronted with the access collisions, the device activity detection becomes a challenging task for the base station (BS) to recognize the sparsely activated devices. In addition, the small-sized data packets in IoT applications also impose new requirement on the transmission efficiency of M2M communications.
\subsection{Related Work}
Different solutions have been proposed to address the problems encountered in crowded RA scenarios. Some grant-based schemes \cite{ACB,seperation,dynamicone,morePA2,reusePA,TVTearly,TVTCARA,NORA} modify the exchange of controlling signals to efficiently allocate resource blocks (RBs) for uplink data transmission while other grant-free schemes exploit the sparseness feature with compressed sensing (CS) algorithms \cite{blockCS,2013CS,CRAN,CS,2015ICC,ICASSP,AMPmassive,AMPmassive2}. In addition, the slotted ALOHA protocol has also been extensively investigated to improve the throughput of RA schemes \cite{Aloha1,Aloha2,fastretrial,fastretrialMTC,CRDSA,IRSA,Capture,CSA,frameless,SCCSA}.
\subsubsection{Grant-Based RA Schemes}
In grant-based RA schemes, each activated device randomly chooses one orthogonal preamble to apply for the corresponding RB for its uplink transmission. These preambles inherently exhibit outstanding detection performances due to their orthogonality. However, the massiveness of M2M communications causes severe overload of the physical random access channel (PRACH) and preamble collisions.

The PRACH overload, i.e., RA congestion when a massive number of active devices apply for a limited number of RBs, further leads to low access probability, high latency and even interruption of service. Some solutions have been proposed for the PRACH overload, such as the Access Class Barring (ACB) scheme \cite{ACB}, delicate splitting of the RA preamble set \cite{seperation} as well as the automatic configuration of the RA channel parameters \cite{dynamicone}. The preamble collision issue is also dealt with from different perspectives. For example, the preamble shortage is addressed either by increasing the number of available preambles \cite{morePA2} or by preamble reusing \cite{reusePA}. Early preamble collision detection schemes were proposed to avoid RB wastage \cite{TVTearly, TVTCARA}. The early collision detection \cite{TVTearly} is performed based on tagged preambles which is also exploited to monitor the RA load. A collision-aware resource access (CARA) scheme was proposed in \cite{TVTCARA} for the efficient use of granted RBs. In addition, a non-orthogonal resource allocation (NORA) \cite{NORA} scheme was proposed to exploit the timing alignment information of devices in different spatial groups.

Unfortunately, a handshaking process is always required by grant-based RA schemes, which causes heavy signaling overhead and undermines the transmission efficiency for small-sized data packets. Furthermore, it is still hard to alleviate the impacts of severe preamble collisions on grant-based RA schemes. As an alternative, some grant-free RA schemes below may serve as promising solutions to crowded M2M communications.
\subsubsection{Compressed Sensing Based Grant-Free RA Schemes}
Several grant-free schemes employ CS algorithms to exploit the sparseness feature of M2M communications. The CS algorithms are performed upon the received pilot signal to accomplish the user activity detection (UAD) and/or channel estimation (CE) problem. For example, a block CS algorithm \cite{blockCS} was proposed for distributed UAD and resource allocation for clustered devices. A greedy algorithm based on orthogonal matching pursuit (OMP) was proposed in \cite{2013CS} for the joint UAD and CE problem. The same task as in \cite{2013CS} is accomplished by a modified Bayesian compressed sensing algorithm \cite{CRAN} for the cloud radio access network (C-RAN). The powerful approximate message passing (AMP) algorithm \cite{CS} was studied for the joint UAD and CE problem when the BS is equipped either with one single antenna \cite{2015ICC,ICASSP} or with multiple antennas \cite{AMPmassive, AMPmassive2}.

The performances of CS-based schemes rely on the \emph{sampling ratio} of CS algorithms. Henceforth, sufficiently long pilot sequences are required due to the massiveness of M2M devices. In addition, pilots are transmitted for the UAD whenever new devices are activated with sporadic data packets, which undermines the transmission efficiency of short packets. Therefore, another kind of grant-free RA schemes, the slotted ALOHA based RA protocols are considered for crowded M2M communications \cite{fastretrial,fastretrialMTC,CRDSA,IRSA,Capture,CSA,frameless,SCCSA}.
\begin{table*}
	\footnotesize
	\renewcommand\arraystretch{1.5}
	\caption{Advantages and disadvantages for different RA schemes.}
	\centering
	\begin{tabular}{ccccccccccccccccccccccc}
		\hline
		RA Scheme&Pros&Cons\\
		\hline
 \multirow{2}{*}{Grant-Based RA scheme} &\multirow{2}{*}{Outstanding detection performance}& Heavy signaling overhead due to handshaking\\   &	& Inevitable access collision\\ 	
		\hline
		CS-Based Grant-Free RA scheme& Capable of joint UAD and CE & Efficiency undermined for small-sized data packets\\
	    \hline
		Slotted ALOHA-Based RA protocols& Higher efficiency due to incorporated data transmission & Vulnerable to collisions\\
		\hline
		\multirow{2}{*}{Fixed-Symbol Aided RA scheme}& Higher efficiency due to incorporated data transmission & \multirow{2}{*}{One symbol sacrificed for device activity detection}\\ & Access collision solved by MUD&\\
		\hline
	\end{tabular}
	\label{procon}
\end{table*}

\subsubsection{Slotted ALOHA Based RA Protocols}
In slotted ALOHA protocols, active devices transmit sporadic data packets in aligned slots while packets in collision-free slots can be correctly recovered. A fast retrial scheme was proposed in \cite{fastretrial} for multi-channel ALOHA. In this scheme, collisions can be immediately recognized and collided packets are re-transmitted in randomly chosen sub-channels in the next slot. The stability of this fast retrial scheme is further analyzed in \cite{fastretrialMTC} with rate control for machine-type communications. For single-channel slotted ALOHA protocols, the throughput is improved by transmitting multiple replicas of the same data packet and employing inter-slot successive interference cancellation (SIC) among replicas. For example, in the contention resolution diversity slotted ALOHA (CRDSA) protocol \cite{CRDSA}, each activated device sends two replicas, i.e., the CRDSA protocol introduces a (2,1) repetition code to the conventional slotted ALOHA protocol. Recently, an irregular repetition slotted ALOHA (IRSA) protocol was proposed in \cite{IRSA} which improves the RA throughput by optimizing the probability for activated devices to choose repetition codes with different rates. This IRSA protocol was extended to the Rayleigh fading channel \cite{Capture}, where collided data packets can still be decoded as long as the signal-to-interference-and-noise ratio (SINR) exceeds a predefined threshold. A coded slotted ALOHA (CSA) scheme \cite{CSA} was proposed employing the general ($n,k$) packet erasure codes to encode the data packets. At the receiver end of the CSA protocol, the maximum-a-posterior decoder and SIC are employed together to recover the collided packets. According to the analogy between the CSA and erasure correcting codes, the frameless IRSA protocol  \cite{frameless} and the spatially coupled RA protocol \cite{SCCSA} were proposed, analogous to rateless codes and spatially coupled LDPC codes, respectively.

Different from CS-based RA schemes, slotted ALOHA based RA protocols are efficient for small-sized data packets since its uplink transmission phase is incorporated in the RA process. Sporadic data packets from newly activated devices can be directly transmitted without another round of UAD at the receiver. However, since the successful decoding of data packets relies heavily on collision-free slots or high SINR, existing schemes such as the IRSA and the CSA could barely work in crowded M2M communications where slots suffer from severe collisions. For reading convenience, the advantages and disadvantages of different RA schemes are summarized in Table \ref{procon}.
\subsection{Contributions}
In order to inherit the advantages of slotted ALOHA based RA protocols and deal with the intra-slot collision, we propose a fixed-symbol aided RA scheme, where active devices access the network in a grant-free manner. In each RA frame, each active device inserts one fixed symbol into its data packet and transmits the packet in one randomly chosen slot. The operations at the BS are divided into two phases. For the first phase, based on the low-complexity message passing algorithms (MPA), an iterative message passing based activity detection (MP-AD) algorithm is proposed to detect the device activity. For the second phase, according to the activity detection result, multi-user detection (MUD) is further employed in each slot to decode the collided data packets.

The MP-AD algorithm is explained as follows. Firstly, we model the fixed-symbol aided RA process by considering the received signals of the fixed symbols in all the slots. According to the system model, a factor graph is established with three different types of nodes. The message update equations for different types of nodes in the factor graph are derived, according to which the BS is enabled to detect the activity of each device in each slot. In order to alleviate the correlation problem of the message passing process in the MP-AD algorithm, we further introduce the deep neural network (DNN) structure. This DNN-aided MP-AD (DNN-MP-AD) algorithm is designed by transforming the edge-type message passing process on a factor graph into a node-type one in a DNN structure. Weights are then imposed on the messages passing in the DNN and further trained to improve the detection accuracy.

Although the activity detection issue is similar to the sparse recovery problem, the proposed MP-AD algorithm differs from existing CS solutions in that the transmission constraint due to the small-sized data packets is considered at every iteration, i.e., only one slot in a RA frame is chosen by each active device to perform data transmission. In addition, the DNN-MP-AD algorithm alleviates the correlation problem by training the weights in the DNN structure. This training process causes no additional online computational complexity since the weights are trained off-line with powerful hardware devices such as the GPU.
\begin{figure*}
\centering
\includegraphics[width=0.85\textwidth]{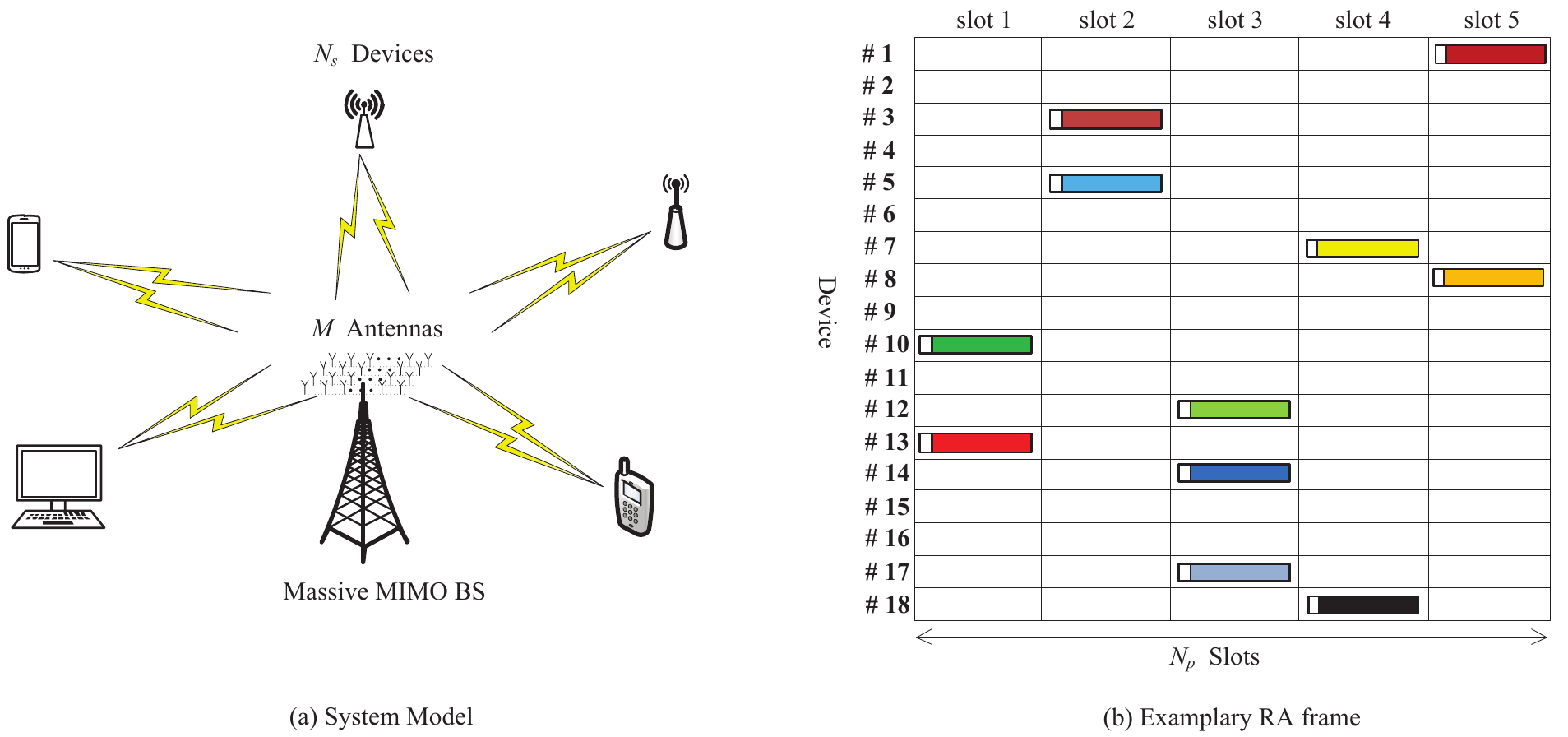}
\caption{System model and one examplary RA frame in the proposed fixed-symbol aided RA scheme. The first fixed symbol in each slot is marked in white while the data symbols of different devices are marked in different colors.}\label{MODEL}
\end{figure*}
The major contributions of this paper are listed as follows.

(i) A fixed-symbol aided RA scheme is proposed for M2M communications. The proposed RA scheme inherits the transmission efficiency of slotted ALOHA protocols and tackles the problem of severe packet collisions via MUD.

(ii) The MP-AD algorithm is designed based on a factor graph where the Bernoulli messages are passed to detect the activity of each device in each slot. This MP-AD algorithm provides essential information for subsequent MUD.

(iii) The DNN-MP-AD algorithm is designed based on the MP-AD framework to alleviate the correlation problem of the messages. By imposing weights on the messages passing in the DNN, the detection accuracy is further improved without causing any additional online computational complexity.
\subsection{Paper Organization}
This paper is organized as follows. The fixed-symbol aided RA scheme is proposed with corresponding system model in Section \ref{sec:model}. According to the system model, a factor graph is presented for the MP-AD algorithm in Section \ref{sec:MPAD}  and the message update equations for different types of nodes are elaborated on. The correlation problem and the DNN-MP-AD algorithm are presented in Section \ref{sec:DNNMPAD} while numerical simulation results are illustrated in Section \ref{sec:simulation}. Finally, the conclusion and future work are given in Section \ref{sec:conclusion}.
\section{Fixed-Symbol Aided Random Access Scheme}\label{sec:model}
\subsection{System Model}\label{smallNp}
As shown in Fig. \ref{MODEL}(a), we consider a M2M communication system with $N_s$ devices centered by a massive MIMO BS. Each device is activated with a certain probability $p_a$, depending on its service type. The BS is equipped with $M$ antennas while each device is with one single antenna. The channel between the devices and the BS is a slow time-varying block fading TDD (time division duplex) channel while the channel state information (CSI) is assumed known to the BS via channel estimation. We consider the application scenarios where devices are placed at fixed locations. Therefore, the coherence time is sufficiently long and the CSI remains constant over a large number of RA frames after one round of channel estimation \cite{stationary,stationary2}. One RA frame is divided into $N_p$ slots. Since the M2M communication in IoT applications is mainly characterized by small-sized data packets, each active device randomly chooses only one slot to perform data packet transmission. Furthermore, the value of $N_p$ is assumed relatively small to reduce the detection latency for activated devices. More details on the system model are discussed at Section \ref{discussion}.

Before transmission, each active device inserts a fixed symbol into its data packet. Without loss of any generality, this fixed symbol is placed at the beginning of the transmitted data packet and its value is fixed as unit value. In order to detect the device activity in each slot, the iterative MP-AD algorithm is performed at the BS upon the received signals of the first symbols in all of the $N_p$ slots. A mathematical model of the received signals is firstly established for the MP-AD algorithm as follows while the details for this algorithm are explained in Section \ref{sec:MPAD}.

Since the  value of the fixed symbol is set as unit value, the received signal on the $l$-th antenna with respect to (w.r.t.) the first symbol in the $t$-th slot can be written as
\begin{equation}\label{model}
y_{lt} = {\sum\limits_{i = 1}^{{N_s}} h_{li}s_{it}}  + n_{lt},
\end{equation}
where $h_{li}$ is the channel gain from the $i$-th device to the $l$-th antenna, $n_{lt}$ is the additive Gaussian noise, and $s_{it}$ is the device-slot indicator, i.e., $s_{ij}=1$ if the $i$-th device selects the $j$-th slot to transmit its data packet. Otherwise, $s_{ij}=0$. The received signal in (\ref{model}) can be rewritten in a matrix form by considering all the slots and all the antennas
\begin{equation}\label{modelmatrix}
{\mathbf{Y}}_{M\times N_p} = { \mathbf{H} }_{M\times N_s}{ \mathbf{S} }_{N_s\times N_p} + {\mathbf{N}}_{M\times N_p},
\end{equation}
where $\mathbf{H}$ is the channel matrix while $\mathbf{N}$ is an additive white Gaussian noise matrix with variance $\sigma_n^2$ for each entry. $\mathbf{S}$ is a device-slot indicator matrix, as well as the target of the MP-AD algorithm and the DNN-MP-AD algorithm.

It is noted that although the problem (\ref{modelmatrix}) is a sparse recovery issue, the target matrix $\mathbf{S}$ follows a different distribution from the ones considered by existing CS algorithms. Since each active device can only choose one slot for transmission while a slot might be chosen by more than one device, each row of the matrix $\mathbf{S}$ has at most one \lq\lq1\rq\rq\ but each column of $\mathbf{S}$ might have more than one \lq\lq1\rq\rq. When a slot $p$ is not chosen by any M2M device, the $p$-th column of matrix $\mathbf{S}$ would be all zeros. Similarly, if a M2M device $s$ is inactive, the $s$-th row of matrix $\mathbf{S}$ would also be all zeros. The probability that a certain row of $\mathbf{S}$ has just one \lq\lq1\rq\rq\ is $p_a$ while the probability for each entry $s_{sp}$ to be \lq\lq1\rq\rq\ is
\begin{equation}\label{p0}
{p_0}=P\left(s_{sp}=1\right)=p_a/N_p.
\end{equation}
\begin{remark}\label{realcomplex}For notational convenience, we consider a real-valued channel matrix for the proposed MP-AD algorithm and the DNN-MP-AD algorithm. Note that this is not in contradiction with the realistic complex-valued channel matrix, since we can regard the complex-valued matrices as a stack of two real-valued matrices, i.e., the following two equations are equivalent
			\begin{equation}\label{complex}
			\left( \mathbf{Y}_R + j\mathbf{Y}_I\right)=\left(\mathbf{H}_R+j\mathbf{H}_I\right)\mathbf{S}+\left(\mathbf{N}_R+j\mathbf{N}_I\right),
			\end{equation}
			\begin{equation}\label{real}
			\left[ \begin{split} {\mathbf{Y}_R}\\{\mathbf{Y}_I} \end{split} \right] = \left[ \begin{split}{\mathbf{H}_R}\\{\mathbf{H}_I} \end{split} \right] \mathbf{S} + \left[ \begin{split} {\mathbf{N}_R}\\{\mathbf{N}_I}\end{split} \right],
			\end{equation}
			where $j^2=-1$, $\mathbf{Y}_R$ and $\mathbf{Y}_I$ are the real part and imaginary part of the complex-valued received matrix, $\mathbf{H}_R$ and $\mathbf{H}_I$ are the real part and imaginary part of the complex-valued channel matrix, and $\mathbf{N}_R$ and $\mathbf{N}_I$ represent real part and imaginary part of the complex noise matrix. Since (\ref{real}) is identical to (\ref{modelmatrix}), the realistic complex-valued problem in (\ref{complex}) can be represented by the real-valued problem (\ref{modelmatrix}). However, it is noted that the transformation from (\ref{complex}) to (\ref{real}) doubles the column number of the matrix representation. Therefore, the $M$ antennas in the MP-AD algorithm actually correspond to $M^*=M/2$ antennas in the realistic complex-valued scenarios.
\end{remark}
\subsection{Random Access Process}
An example of one RA frame in the proposed fixed-symbol aided RA scheme is shown in Fig. \ref{MODEL}(b). In the data packet transmitted in one slot, the first symbol marked in white represents the fixed symbol which takes the unit value while the following symbols marked in different colors represent the data symbols of different devices.

At the receiver end, the received signals with respect to the first symbols in all of the $N_p$ slots are processed by the MP-AD algorithm, according to which, the BS is enabled to detect the activity of each device in each slot. The received signals with respect to the data symbols in each slot are firstly stored in the memory. After the device activity detection, the BS performs subsequent MUD in each slot to decode the collided data packets. For example, if the device activity is correctly detected in slot 2, the subsequent MUD is only performed for 2 active devices, i.e., device 3 and device 5 while inactive devices in this slot are ignored.

There are many off-the-shelf MUD algorithms which can be directly applied to the massive MIMO BS. For example, the low-complexity Gaussian message passing iterative detector (GMPID) \cite{GBPMUD,Lei2016Convergence,Lei2016Gaussian} can be employed for MUD when the data symbols are Gaussian distributed, which is discussed at Section \ref{discussion}. In addition, due to the low activation probability $p_a$, the number of collided packets in each slot is much smaller than $N_s$. As a result, excellent MUD performance can be expected as long as the device activity is correctly detected by the MP-AD algorithm in the corresponding slot.

The proposed fixed-symbol aided RA scheme inherits the uplink transmission efficiency of slotted ALOHA protocols and only one fixed symbol is sacrificed for the device activity detection. The severe intra-slot collision issue in crowded M2M communications is solved by the MUD, which is enabled by the device activity detection result of the following MP-AD algorithm. The proposed fixed-symbol aided random access scheme is summarized in Scheme \ref{alg} while the MP-AD algorithm is explained as follows.
\section{Message Passing Based Activity Detection Algorithm}\label{sec:MPAD}
\begin{figure}
	\centering
          \includegraphics[width=0.48\textwidth]{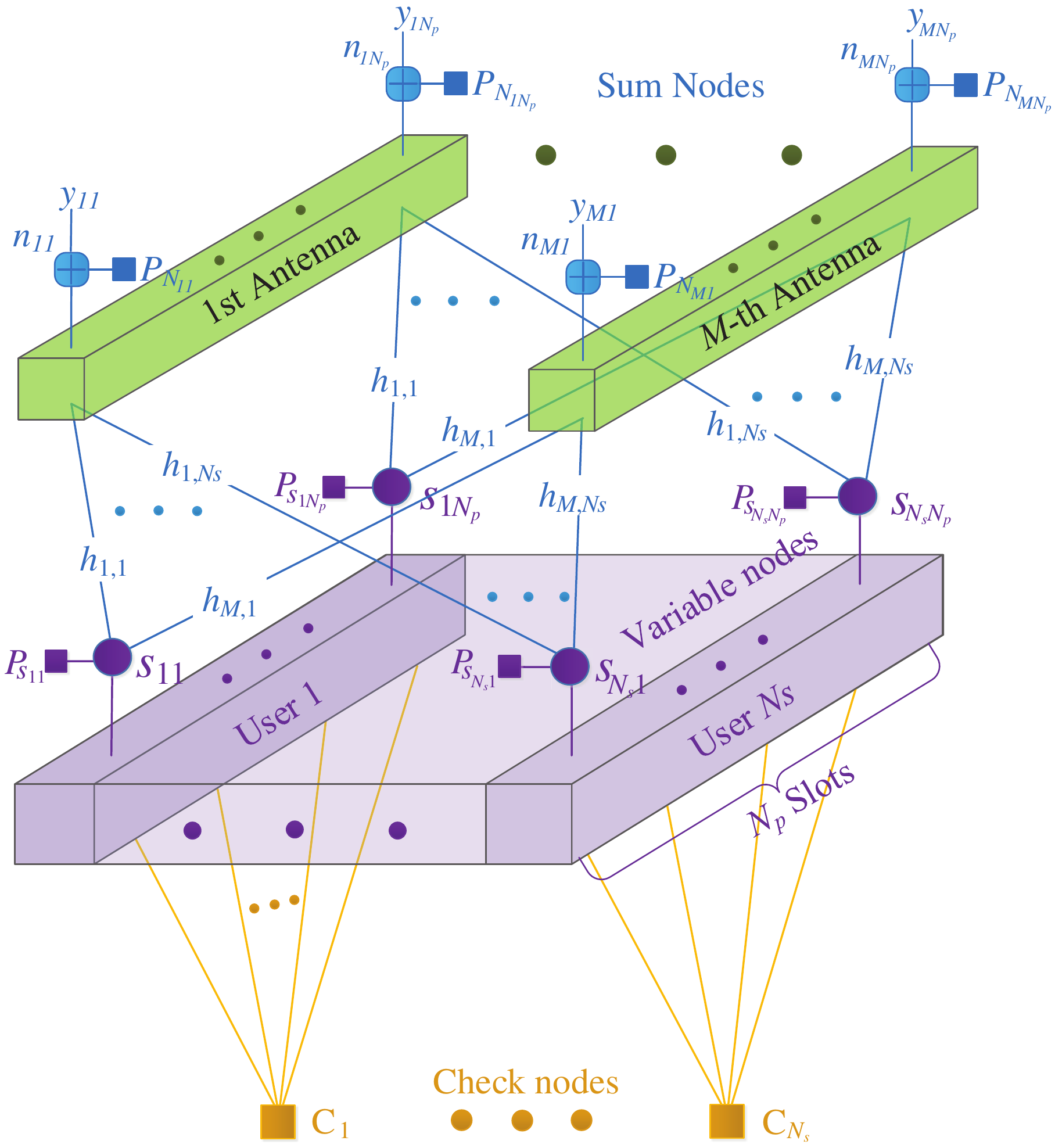}
\caption{Factor graph of the MP-AD algorithm.}\label{factorgraph}
\end{figure}
\newcounter{TempEqCnt}
\setcounter{TempEqCnt}{\value{equation}}
\setcounter{equation}{7}	
\begin{figure*}
	\begin{equation}\label{sum_Ber}
	\begin{split}
	p_{mp \to sp}^s(\tau )&=\frac{{P\left( {{s_{sp}} = 1|{y_{mp}, {\mathbf{h}}_m, {\mathbf{p}}^{vs}_{mp,\sim sp}(\tau)}} \right)}}{{P\left( {{s_{sp}} = 1|{y_{mp}, {\mathbf{h}}_m, {\mathbf{p}}^{vs}_{mp,\sim sp}(\tau)}} \right) + P\left( {{s_{sp}} = 0|{y_{mp}, {\mathbf{h}}_m, {\mathbf{p}}^{vs}_{mp,\sim sp}(\tau)}} \right)}}\\
	&=\frac{{f\left( {{y_{mp}}|u_{mps}^*(\tau )+ {h_{ms}} ,v_{mps}^*(\tau )} \right)}}{{f\left( {{y_{mp}}|u_{mps}^*(\tau )+ {h_{ms}},v_{mps}^*(\tau )} \right) + f\left( {{y_{mp}}|u_{mps}^*(\tau ),v_{mps}^*(\tau )} \right)}}\\
	&={\left[ {{\rm{1 + exp}}\left( {\dfrac{{h_{ms}^2 - 2{h_{ms}}({y_{mp}} - u_{mps}^*(\tau ))}}{{2v_{mps}^*(\tau )}}} \right)} \right]^{ - 1}},
	\end{split}
	\end{equation}\hrulefill
\end{figure*}
\setcounter{equation}{\value{TempEqCnt}}
The message passing algorithm (MPA) is renowned for its feasible implementation complexity since the overall processing can be departed into distributed calculations, which is suitable for parallel execution. As a result, the MPA has been widely applied for Compressed Sensing \cite{CS}, MUD \cite{GBPMUD,Lei2016Convergence,Lei2016Gaussian}, channel estimation \cite{GMPCW} and the Belief Propagation (BP) decoding for LDPC codes \cite{William2009}. Therefore, it is practical for the powerful massive MIMO BS to perform the iterative message passing algorithm to detect the activity of each device in each slot. The factor graph for the MP-AD algorithm is presented as follows.
\subsection{Factor Graph Representation}
The system model described in Section \ref{sec:model} can be represented by a factor graph in Fig. \ref{factorgraph} and the messages passing on the graph are the likelihood messages for the Bernoulli variables, i.e., the entries in matrix $\mathbf{S}$. As shown in Fig.  \ref{factorgraph}, there are three types of nodes in the factor graph, i.e., sum nodes (SNs), variable nodes (VNs) and check nodes (CNs). Each SN  $y_{mp}$ stands for an entry in matrix $\mathbf{Y}$, i.e., the received signal on the $m$-th antenna with respect to the first symbol in the $p$-th slot. Each VN $s_{sp}$ is a Bernoulli variable and represents the $p$-th element of the $s$-th row in matrix $\mathbf{S}$. Furthermore, the CNs stand for the check node constraints for the devices, i.e., at most one slot can be chosen by a device in a RA frame.

According to the factor graph, the message updating diagram among SNs, VNs and CNs is illustrated in Fig. \ref{updatediagram}. According to the principle of the MPA, the output message, defined as the extrinsic information, is derived by the incoming messages from the other edges that are connected to the same node. The message update equations for the SNs, VNs and CNs are derived as follows.
\subsection{Message Update at Sum Nodes}
\begin{figure*}
	\centering
	\includegraphics[width=0.7\textwidth]{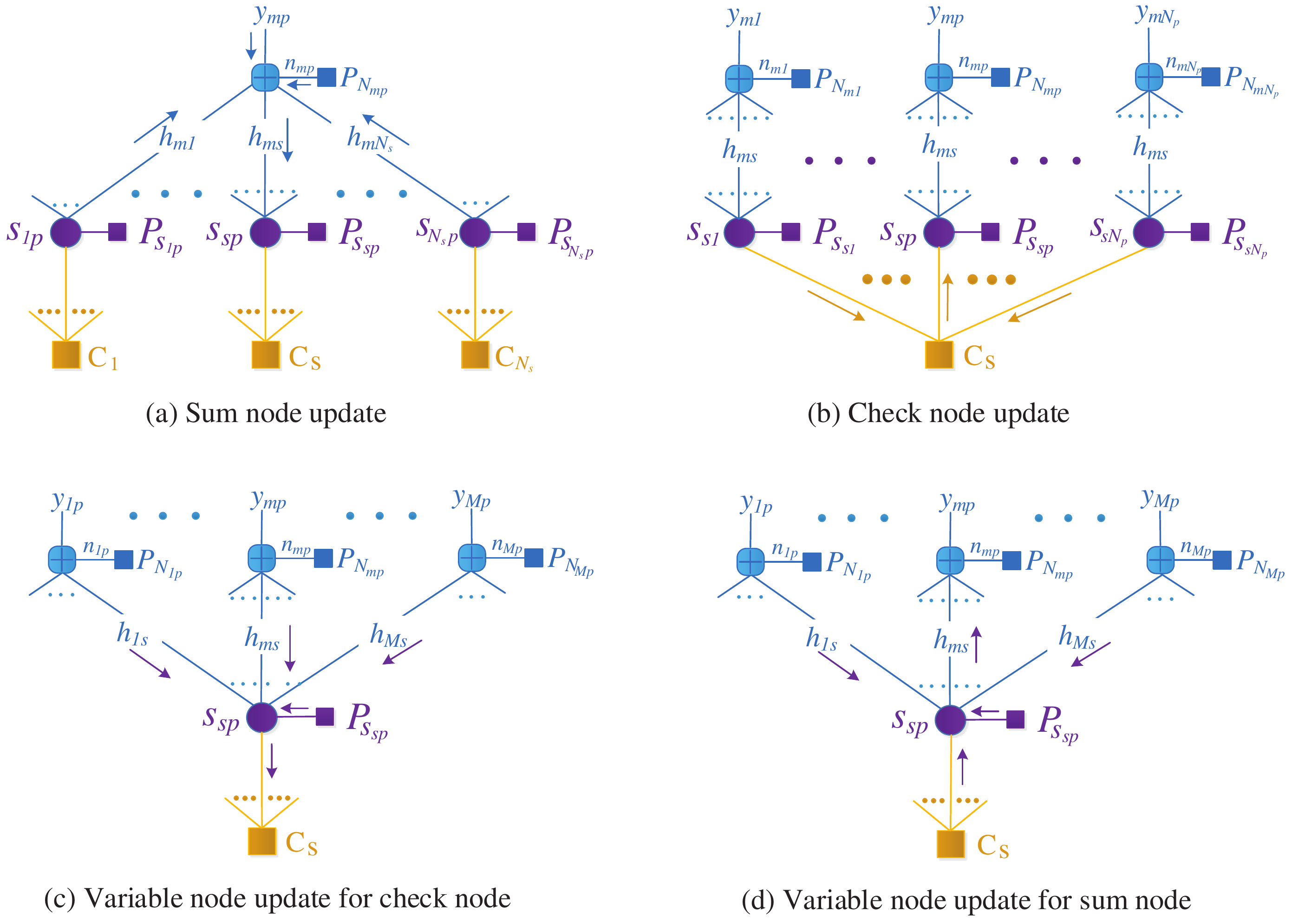}\\
	\caption{Message update at the sum nodes, check nodes and variable nodes. The messages passing on the edges are the non-zeros probabilities or LLR of the Bernoulli variables in the indicator matrix $\mathbf{S}$. The output message called extrinsic information is derived by the messages on the other edges that are connected with the same node.}\label{updatediagram}\vspace{-0.3cm}
\end{figure*}

The message update at each SN can be seen as a multiple-access process and the extrinsic message from the $mp$-th SN to the $sp$-th VN is presented as an example in Fig. \ref{updatediagram}(a). Firstly, the received signal $y_{mp}$ at the $mp$-th SN can be rewritten to
\begin{equation}\label{expandy}
\begin{split}
y_{mp}&={h_{ms}s_{sp}} + \sum\limits_{i \in {{\cal N}_s}/s}{{h_{mi}}{s_{ip}}}  + {n_{mp}}\\
&={h_{ms}s_{sp}} +n^*_{mps}(\tau),
\end{split}
\end{equation}
where $s\in \mathcal{N}_s$ for $\mathcal{N}_s =\{1,\ldots,N_s\}$, $m\in\mathcal{M}$ for $\mathcal{M}=\{1,\ldots, M\}$ and $p\in\mathcal{N}_p$ for $\mathcal{N}_p=\{1,\ldots,N_p\}$. We assume that $p_{sp\to mp}^{vs}(\tau)$ denotes the non-zero probability for the Bernoulli variable $s_{sp}$ passing from the $sp$-th VN to the $mp$-th SN in the $\tau$-th iteration. According to the incoming messages from the VNs and the \emph{central limit theorem}, the SN approximates $n_{mps}^*(\tau)$ as an equivalent Gaussian noise with mean $u_{mps}^*(\tau)$ and variance $v_{mps}^*(\tau)$,
\begin{equation}\label{equ_noi}
\begin{split}
u_{mps}^*(\tau)&=\sum\limits_{i\in {\cal N}_s/s} {h_{mi}}p_{ip \to mp}^{vs}(\tau),\\
v_{mps}^*(\tau)&= \sum\limits_{i\in {\cal N}_s/s} {h_{mi}^2p_{ip \to mp}^{vs}(\tau)q_{ip \to mp}^{vs}(\tau)+\sigma _n^2},
\end{split}
\end{equation}
where $p_{ip\to mp}^{vs}(\tau)$ and $q_{ip \to mp}^{vs}(\tau)$ are the non-zero and zero probabilities for the Bernoulli variable $s_{ip}$ passing from the $ip$-th VN to the $mp$-th SN. Then the extrinsic message $p_{mp\to sp}^s(\tau)$ from the $mp$-th SN to the $sp$-th VN is derived in (\ref{sum_Ber}),	where $\mathbf{h}_m$ is the $m$-th row of $\mathbf{H}$, ${\mathbf{p}}_{mp,\sim sp}^{vs}(\tau)$ is the set of Bernoulli probabilities $\{{{p}}_{ip\to mp}^{vs}(\tau)|i\in \mathcal{N}_s/s\}$ and $f(x|u,v)$ is the Gaussian distribution \emph{probability density function} for $x$,
\begin{equation}\nonumber
f\left(x|u,v \right)=\dfrac{1}{\sqrt {2\pi v}}e^{-\dfrac{(x-u)^2}{2v}}.
\end{equation}
\setcounter{equation}{8}

To avoid overflow caused by a large number of multiplications of probabilities, the Bernoulli messages are represented in the form of \emph{log-likelihood ratio} (LLR) instead of the non-zero probability in (\ref{sum_Ber}). For example, the LLR message $l_{mp \to sp}^s (\tau)$ passing from the $mp$-th SN to the $sp$-th VN in the $\tau$-th iteration is derived by performing the logarithmic operation on the ratio of the non-zero probability $p_{mp \to sp}^s(\tau )$ to the zero probability $q_{mp \to sp}^s(\tau )$
\begin{equation}\label{SUMNODE}
l_{mp \to sp}^s (\!\tau\!)\! =\! \log \dfrac{{p_{mp \to sp}^s(\!\tau\!)}}{{q_{mp \to sp}^s(\!\tau\!)}}\overset{(a)}{\!=\!}{\dfrac{2(y_{mp}\!-\!u_{mps}^* (\!\tau\!))h_{ms}\!-\!h_{ms}^2}{2v_{mps}^*(\!\tau\!)}}
\end{equation} where ($a$) is derived by substituting the result of (\ref{sum_Ber}) into the derivation of $l_{mp \to sp}^s (\tau)$ as well as the fact that $q_{mp \to sp}^s(\tau ) + p_{mp \to sp}^s(\tau ) =1$ for each Bernoulli variable $s_{sp}$.
\subsection{Message Update at Variable Nodes}
The message update at each VN can be seen as a broadcasting process and the extrinsic message from the VN is derived following the message combination rule \cite{Loeliger2006}, i.e., the extrinsic message is a normalized product of the input probabilities.
\subsubsection{Message update for sum nodes}
The message update from the $sp$-th VN to the $mp$-th SN is presented as an example in Fig. \ref{updatediagram}(d). The extrinsic message is derived by the initial probability $p_0$ of each VN, the incoming message $p_{s\to sp}^c(\tau )$ from the $s$-th CN and the incoming messages $p^s_{jp\to sp}(\tau)$ from the $jp$-th SN with $j\in\mathcal{M}/m$. According to the message combination rule, the extrinsic message from the $sp$-th VN to the $mp$-th SN is
\begin{equation}\label{var_ber1}
\begin{split}
&p_{sp \to mp}^{vs}(\tau \!+\!1)\!=\! P({s_{sp}\!=\!1}|{\mathbf{p}}_{sp,\sim mp}^s(\tau ),p_{s \to sp}^c(\tau ),{p_0})\\
&\overset{(b)}{=}\!\! \dfrac{{p_0}p_{s \to sp}^c(\tau )\! \! \!\! \prod\limits_{\! j \in {\cal M}/m} {\!\!\!\! p_{jp \to sp}^s}(\tau )}{{p_0}p_{s \to sp}^c(\tau )\! \! \!\! \prod\limits_{\! j \in {\cal M}/m} {\!\!\!\! p_{jp \to sp}^s}(\tau )\!+\!{q_0}q_{s \to sp}^c(\tau )\! \! \!\! \prod\limits_{\! j \in {\cal M}/m} {\!\!\!\! q_{jp \to sp}^s}(\tau )}
\end{split}
\end{equation}
where ${\mathbf{p}}_{sp,\sim mp}^s(\tau)=\{p^s_{jp\to sp}(\tau)|j\in\mathcal{M}/m\}$ and ($b$) is derived by the normalized product of the input probabilities.
\subsubsection{Message update for check nodes}
Similarly, in Fig. \ref{updatediagram}(c), the extrinsic message update from the $sp$-th VN to the $s$-th CN is derived by the initial probability ${p}_{0}$ and the incoming messages ${p_{mp \to sp}^s} (\tau )$ from the $mp$-th SN with $m\in\mathcal{M}$.
\begin{equation}\label{var_ber2}
\begin{split}
&p_{sp \to s}^{vc}(\tau  + 1)= P({s_{sp}=1}|{\mathbf{p}}_{sp}^s(\tau ),{p_0})\\
&\overset{(c)}{=} \dfrac{   {p_0}\prod\limits_{m \in {\cal M}} {p_{mp \to sp}^s}(\tau )   }{    {p_0}\prod\limits_{m \in {\cal M}} {p_{mp \to sp}^s}(\tau ) + {q_0}\prod\limits_{m \in {\cal M}} {q_{mp \to sp}^s}(\tau ) },
\end{split}
\end{equation}
where ${\mathbf{p}}_{sp}^s(\tau)=\{p_{mp\to sp}^s(\tau)|m\in\mathcal{M}\}$ and ($c$) is derived by the normalized product of the input Bernoulli probabilities.

Analogous to (\ref{SUMNODE}), the extrinsic LLR messages from the VNs are derived as follows
\begin{equation}\label{SDvnupdate}
\begin{split}
l_{sp \to mp}^{vs}(\tau\!+\! 1)&\!=\!\log \dfrac{{p_{sp \to mp}^{vs}(\tau \!+\! 1)}}{{q_{sp \to mp}^{vs}(\tau \!+\! 1)}}\\
&\!=\!\sum\limits_{j \in \mathcal{M}/m} {l_{jp \to sp}^s(\tau )}  + {{\bar l}_{sp}} + l_{s \to sp}^c(\tau ),\\
l_{sp \to s}^{vc}(\tau  \!+\! 1)&\!=\!\log \dfrac{p_{sp \to s}^{vc}(\tau  \!+\! 1)}{q_{sp \to s}^{vc}(\tau  \!+\! 1)}\!=\!\! \sum\limits_{m \in \mathcal{M}} {l_{mp \to sp}^s(\tau )}  \!+\! {{\bar l}_{sp}},
\end{split}
\end{equation}
where ${\bar l}_{sp}=\log({p}_{0}/{q}_{0})$ and $l_{s\to sp}^c(\tau)$ is defined as follows in (\ref{check_update_LLR}) while $l_{mp\to sp}^s(\tau)$ and $l_{jp\to sp}^s(\tau)$ are derived in (\ref{SUMNODE}).
\subsection{Message Update at Check Nodes}
The $s$-th CN represents a constraint for the corresponding VNs that a VN $s_{sp}=1$ if and only if the $s$-th device is active and  the other VNs $s_{sk}=0$ for any $k\in \mathcal{N}_p/p$. As illustrated in Fig. \ref{updatediagram}(b), the extrinsic message from the $s$-th CN to the $sp$-th VN is derived by the initial activation probability $p_a$ for this device and the incoming messages from the $sk$-th VN with $k\in \mathcal{N}_p/p$. The message update is presented as
\begin{equation}\label{check_update}
\begin{split}
p_{s \to sp}^{c}(\tau)&= P\big(s_{sp}=1|p_a,\mathbf{p}_{s,\sim p}^{vc}(\tau)\big)\\
&={p_a}\mathop \prod \limits_{k\in \mathcal{N}_p/p} (1-p_{sk\to s}^{vc}(\tau)),
\end{split}
\end{equation}
where $\mathbf{p}_{s,\sim p}^{vc}(\tau)=\{{{p}}_{sk\to s}^{vc}(\tau)|k\in \mathcal{N}_p/p\}$. The extrinsic LLR message $l_{s\to sp}^c(\tau)$ passing from the $s$-th CN to the $sp$-th VN is derived by
\begin{equation}\label{check_update_LLR}
l_{s \to sp}^{c}(\tau)= \log \dfrac{p_{s \to sp}^{c}(\tau)}{q_{s \to sp}^{c}(\tau)}=-\log\big( e^{-\tilde{l}_{s \to sp}^{c}(\tau)} - 1\big),
\end{equation}
where
\begin{equation}\label{CN2VN}
\begin{split}
\tilde l_{s \to sp}^c(\tau)&=\log\left( {p_{s \to sp}^c(\tau )} \right)\\
&=\log ({p_a})-\sum\limits_{k \in {{\cal N}_p}/p}{\log (} {e^{l_{sk \to s}^{vc}(\tau )}} + 1).
\end{split}
\end{equation}
\subsection{Output and Decision}
The final output Bernoulli message for each VN is derived by all the incoming messages from SNs and the CN as well as the initial probability $p_0$,
\begin{equation}\label{var_dec}
\hat{p}_{sp}(\!\tau\!)\!\!=\!\!\dfrac{ {p_0}p_{s\to sp}^c(\!\tau\!) \!\!\!\!\mathop \prod \limits_{m \in \mathcal{M}} \!\!\!p_{mp \to sp}^s(\!\tau\!) }{{p_0}p_{s\to sp}^c(\!\tau\!) \!\!\!\!\mathop \prod \limits_{m \in \mathcal{M}} \!\!\!p_{mp \to sp}^s(\!\tau\!) \!+\! {q_0}q_{s\to sp}^c(\!\tau\!) \!\!\!\!\mathop \prod \limits_{m \in \mathcal{M}}\!\!\!q_{mp \to sp}^s(\!\tau\!)}
\end{equation}
The output LLR message for each Bernoulli variable $s_{sp}$ is
\begin{equation}\label{var_dec_LLR}
\hat{l}_{sp}(\tau) =\log \dfrac{\hat{p}_{sp}(\tau)}{\hat{q}_{sp}(\tau)} =\mathop \sum \limits_{m \in \mathcal{M}} l_{mp \to sp}^s(\tau)+ \bar{l}_{sp} + l_{s\to sp}^c(\tau).
\end{equation}
Then, the decision for each Bernoulli variable is
\begin{equation}\label{zerone}
\hat{s}_{sp} =\left\{ \begin{split}
1, \;\;\mathrm{if}\;\; \hat{l}_{sp}\geq0 \\
0, \;\;\mathrm{if}\;\; \hat{l}_{sp}<0
\end{split} \right.
\end{equation}
Finally, $\widehat{\mathbf{S}}\!=\![\hat{{s}}_{sp}]_{N_s\times N_p}$ is the output of the MP-AD algorithm. For reading convenience, the MP-AD algorithm is summarized in Phase 1 of Scheme \ref{alg}. For Phase 2 (MUD phase) of Scheme \ref{alg}, the sub-matrix $\mathbf{H_p}$ is obtained according to $\widehat{\mathbf{S}}$, i.e., $\mathbf{H_p}$ is composed of the columns of $\mathbf{H}$ corresponding to the active devices in slot $p$.
\begin{algorithm}[t]
	\renewcommand{\algorithmcfname}{Scheme}
	\!\!\!\textbf{Phase 1:} \emph{Device Activity Detection}
	
	\KwIn{received matrix $\mathbf{Y}$ w.r.t. fixed symbols in (\ref{modelmatrix}); $\mathbf{H}$, $p_a$, $p_0$, maximum iteration number $L$}
	
	\KwOut{Detected indicator matrix $\widehat{\mathbf{S}}\!=\![\hat{{s}}_{sp}]_{N_s\times N_p}$}
	\textbf{Initialization:} $\tau=0$, $l^{vs}_{sp\to mp}(\tau)=0$, $\forall s\in \mathcal{N}_s,p\in \mathcal{N}_p,m\in \mathcal{M}$;
	
	\For{ $\tau=1,...,L$ and $\forall s\in \mathcal{N}_s,p\in \mathcal{N}_p,m\in \mathcal{M}$}
	{		
		SN update: update $l^s_{mp\to sp}(\tau)$ by (\ref{SUMNODE});
		
		VN update for CN: update $l^{vc}_{sp\to s}(\tau)$ by (\ref{SDvnupdate});
		
		CN update: update $l^{c}_{s\to sp}(\tau)$ by(\ref{check_update_LLR});
		
		VN update for SN: update $l^{vs}_{sp\to mp}(\tau)$ by (\ref{SDvnupdate});
		
		$\tau=\tau+1$;
	}
	
	\textbf{Output message:} calculate the output message (\ref{var_dec_LLR});
	
	\textbf{Hard decision:} make the hard decision by (\ref{zerone});
	
	\KwResult{$\widehat{\mathbf{S}}\!=\![\hat{{s}}_{sp}]_{N_s\times N_p}$}

	\!\!\!\textbf{Phase 2:} \emph{Multi-User Detection}
	
	\KwIn{$\widehat{\mathbf{S}}\!=\![\hat{{s}}_{sp}]_{N_s\times N_p}$, $\mathbf{H}$, received data signals $\mathbf{Y_p}$ stored in buffer for slot $p=1,\ldots,N_p$.}
	
	\KwOut{Recovered data for active devices in the current RA frame.}
	\For{ $p=1,...,N_p$}
	{According to $\mathbf{Y_p}$ and $\mathbf{H_p}$, recover data of active
		
		devices with GMPID\cite{Lei2016Convergence} in slot $p$.
	}		
	\caption{Fixed-Symbol Aided RA Scheme}
	\label{alg}
\end{algorithm}

\subsection{Discussions}\label{discussion}
\subsubsection{Channel State Information} Normally, the channel matrix is affected by both the large-scale fading and the small-scale fading. Since we consider devices with relatively fixed locations, the large-scale fading is assumed constant and can be compensated by the \emph{reverse power control}, i.e., each device adjusts its transmission power to guarantee approximately identical mean received power for all the devices at the BS. Therefore, the channel matrix $\mathbf{H}$ is only characterized by the small-scale fading. Then, the devices transmit pilot sequences to facilitate the uplink channel estimation at the BS.

It is noted that for IoT application scenarios, such as the Phase Management Unit (PMU) in smart grid \cite{PMU,smartgrid} and energy management system in smart homes \cite{energymanagement,smarthome}, the devices are stationary or quasi-stationary and the coherence time of the uplink transmission is therefore sufficiently long \cite{stationary,stationary2}. For example, the experimental results in \cite{stationary} show that in a static testing environment, the channel is constant within tens of minutes and the channel response remains a strong time correlation even under human interference. In addition, as shown by simulations in Section \ref{variation}, the MP-AD algorithm and the DNN-MP-AD algorithm exhibit tolerance against the channel variation. Therefore, after one round of channel estimation, the uplink channel is assumed constant over different RA frames.

\subsubsection{Synchronization}  As stated above, the M2M devices are assumed stationary with fixed locations in this paper. As a result, the transmission delay from each device to the BS can be acquired when it is registered in the network. In this way, synchronization among all the devices can be performed by exploiting the timing advance information. Therefore, we assume that the data packets from the active devices are symbol-wise synchronized.

\subsubsection{Subsequent MUD} The subsequent MUD is performed according to the activity detection result of the MP-AD algorithm. It is noted that the number of collided packets in each slot is much smaller than the number of devices, which greatly reduces the computational complexity of MUD at the BS, especially when the activation probability is low. There are many off-the-shelf MUD algorithms which can be directly applied to the massive MIMO BS. For example, when the data symbols are assumed Gaussian distributed, the performance of the GMPID is proven to converge to that of the minimum mean square error (MMSE) estimator \cite{Lei2016Convergence}. Furthermore, aided by a 10-bit superposition coded modulation scheme \cite{CodedModulation1,CodedModulation2}, the GMPID can still guarantee excellent decoding accuracy for discrete data symbols. Therefore, the accuracy of the device activity detection is the key to the throughput of the proposed fixed-symbol aided RA scheme.

\subsubsection{Energy Consumption and Storage Overhead}
As shown in Fig. \ref{simulation}, the proposed MP-AD algorithm exhibits outstanding detection accuracy in the low-SNR regime. As a result, the transmitting power can be effectively lowered in the proposed fixed-symbol aided RA scheme and the energy consumption is therefore feasible for the devices. At the receiver end, since the message passing algorithm is employed for both the MP-AD algorithm and subsequent MUD, the massive MIMO BS can depart the overall processing into parallel-executed distributed computations. Due to the low computational complexity, the energy consumption also remains feasible for the BS. Therefore, compared with the CS-based grant-free RA schemes and slotted-ALOHA based RA protocols, no extra energy is sacrificed in the proposed RA scheme. In addition, compared with grant-based RA schemes, the proposed RA scheme can greatly reduce the energy consumption of the devices since no handshaking process is required.

In SIC-based slotted ALOHA protocols, collisions needs to be stored to facilitate subsequent inter-slot SIC. However, this storage overhead can be excessively high due to the severe access collisions in crowded M2M communications. By contrast, both the CS-based RA scheme and the proposed fixed-symbol aided RA scheme only need to store the received signals in the current RA frame. As mentioned in Section \ref{smallNp}, the number of slots $N_p$ in one RA frame is assumed relatively small to reduce the processing delay. In addition, the data packets in each slot are normally small-sized for IoT applications. Therefore, the storage overhead also remains feasible for the proposed RA scheme.

\section{DNN-Aided Message Passing Based Activity Detection Algorithm}\label{sec:DNNMPAD}
\subsection{Correlation Problem}\label{correlation}
The message update in message passing algorithms is derived based on the assumption that the messages are mutually independent. However, for the MP-AD algorithm, the messages suffer from the correlation problem, which undermines the accuracy of device activity detection. The correlation problem is caused by the following reasons.

According to the factor graph in Fig. \ref{factorgraph}, the connection between SNs and VNs is characterized by many short cycles with girth 4. As a result, the messages passing on the factor graph can be strongly correlated. In addition, the correlation problem is also caused by the CN update. Rewrite the update equation (\ref{check_update_LLR}) for the $s$-th CN as follows,
\begin{equation}\label{CNinOne}
l_{s \to sp}^{c}(\tau)=-\log\big( e^{\sum\limits_{k \in {{\mathcal N}_p}/p}{\log (} {e^{l_{sk \to s}^{vc}(\tau )}} + 1)-\log ({p_a})} - 1\big).
\end{equation}
Assuming $c$ as a constant, we can observe that the function $\log\big(e^l+c\big)$ is dominated by $c$ when $l$ is negative with a large absolute value. On the contrary, this function is dominated by $l$ and can be approximated as $\log\big(e^l+c\big)\approx l$ if $l$ is a large positive number. We assume that in the accumulative summation term of (\ref{CNinOne}), the message $l_{sk^+ \to s}^{vc}(\tau )$ is a large positive number while the other messages $l_{sk \to s}^{vc}(\tau)$ for $k\in{{{\mathcal N}_p}/\{p,k^+\}}$ are negative or positive but with limited absolute value. Then the message $l_{s \to sp}^{c}(\tau)$ in (\ref{CNinOne}) is approximately proportional to $-l_{sk^+ \to s}^{vc}(\tau )$. This can be explained by the fact that if a VN $s_{sk^+}$ is highly likely to be 1, then the other VNs $s_{sp}$ for $p\in\mathcal{N}_p/k^+$ are highly likely to be 0.  Since $l_{s \to sp}^{c}(\tau)$ is passed to VN $s_{sp}$, the extrinsic message from VN $s_{sp}$ to SN is strongly correlated with the extrinsic message from VN $s_{sk^+}$ to SN. This correlation can be helpful for the iterative convergence when the LLR message $l_{sk^+ \to s}^{vc}(\tau )$ is reliable. However, when false alarm occurs, $l_{sk^+ \to s}^{vc}(\tau)$ is a large positive number for VN $s_{sk^+}=0$. Then the correlation problem of the CN update will cause error propagation on the entire factor graph. According to \cite{Lei2016Convergence}, the extrinsic messages from VNs to CNs are more likely to be unreliable in overloaded MIMO systems where the antennas fail to sustain a large number of devices. In addition, the correlation problem in overloaded systems is even worsened when the SNR is high since the absolute value of the unreliable LLR message is larger with higher SNR.

The correlation problem encountered by the GMPID in \cite{Lei2016Convergence} is solved by a scale-and-add (SA) method where the SA-GMPID is derived by analyzing the convergence of the message passing process. However, the correlation problem caused by the CN update \label{rewriteCN} greatly complicates the theoretical analysis for the proposed MP-AD algorithm. As an alternative, the emerging deep learning method has been proven effective in improving the performance of BP decoders on densely connected Tanner graphs \cite{DNN-BP}. Motivated by this result, we resort to the weighted message passing in DNN and transform the edge-type message passing process in Fig. \ref{updatediagram} into a node-type one in a DNN structure. After the transformation, weights are imposed on the messages in the DNN and further trained to alleviate this correlation problem.
\subsection{Preliminaries to Neural Networks}
In order to make this work self-contained, we provide some preliminaries on the neural networks \cite{DNN-origin}. In terms of the topology, the neural network has several layers of nodes (neurons) while the first layer is termed the \emph{input layer}, the last layer is termed the \emph{output layer} and the remaining layers in the middle are \emph{hidden layers}. Typically, for each node in a hidden layer of the neural network, its input includes the incoming messages from the neighboring nodes in the previous layer and a bias term. The output of this hidden node is obtained by the summation of the input messages, followed by a non-linear activation function, e.g., the $sigmoid$ function to introduce non-linearity. In addition, the nodes in the hidden layers are usually fully or densely connected to the neighbors in the previous layer. In the training phase, weighting parameters are assigned to the messages passing in the neural network while the \emph{loss function} (e.g. the \emph{cross entropy} function or the \emph{Mean Square Error} function) is employed to measure the distance between the output and the true values (i.e., the correct output). Then, the weights are trained with the samples in the \emph{training set} to minimize this loss function. Mathematically, the neural network can be considered as a function of the input neurons and the weights in a hyperspace. With a set of well-trained weights, the neural network can serve as a \emph{classifier} which produces the correct output of future input.

The DNN, one fundamental framework in deep learning, features a large number of hidden layers. Recently, the emerging DNN has been proven effective in a wide variety of communication scenarios such as MIMO communication\cite{deep-MIMO-detection, DLBMIMOC} and sparse vector recovery\cite{OptimalDNNLap}. For the decoder design of channel codes, a DNN-based BP decoder was proposed for short BCH codes \cite{DNN-BP} to alleviate the influence of short cycles in densely connected Tanner graphs.
\subsection{Neural Network Structure for DNN-MP-AD Algorithm}
Different from the typical neural network, the neural network structure for the DNN-MP-AD algorithm is constructed by directly transforming the edge-type message passing on the factor graph into a node-type one, i.e., each edge in the factor graph is now transformed into a node in the hidden layer. Therefore, every hidden layer represents one specific message update in Fig. \ref{updatediagram} and different from the typical fully-connected neural network, the inter-layer connection for the DNN-MP-AD algorithm is uniquely determined by the factor graph in Fig. \ref{factorgraph}. Since the messages passing in the DNN-MP-AD algorithm are LLR messages, no non-linear activation function is considered and the bias term is determined by the message update equations in the MP-AD algorithm.

\begin{figure*}
	\centering
	\includegraphics[width=1\textwidth]{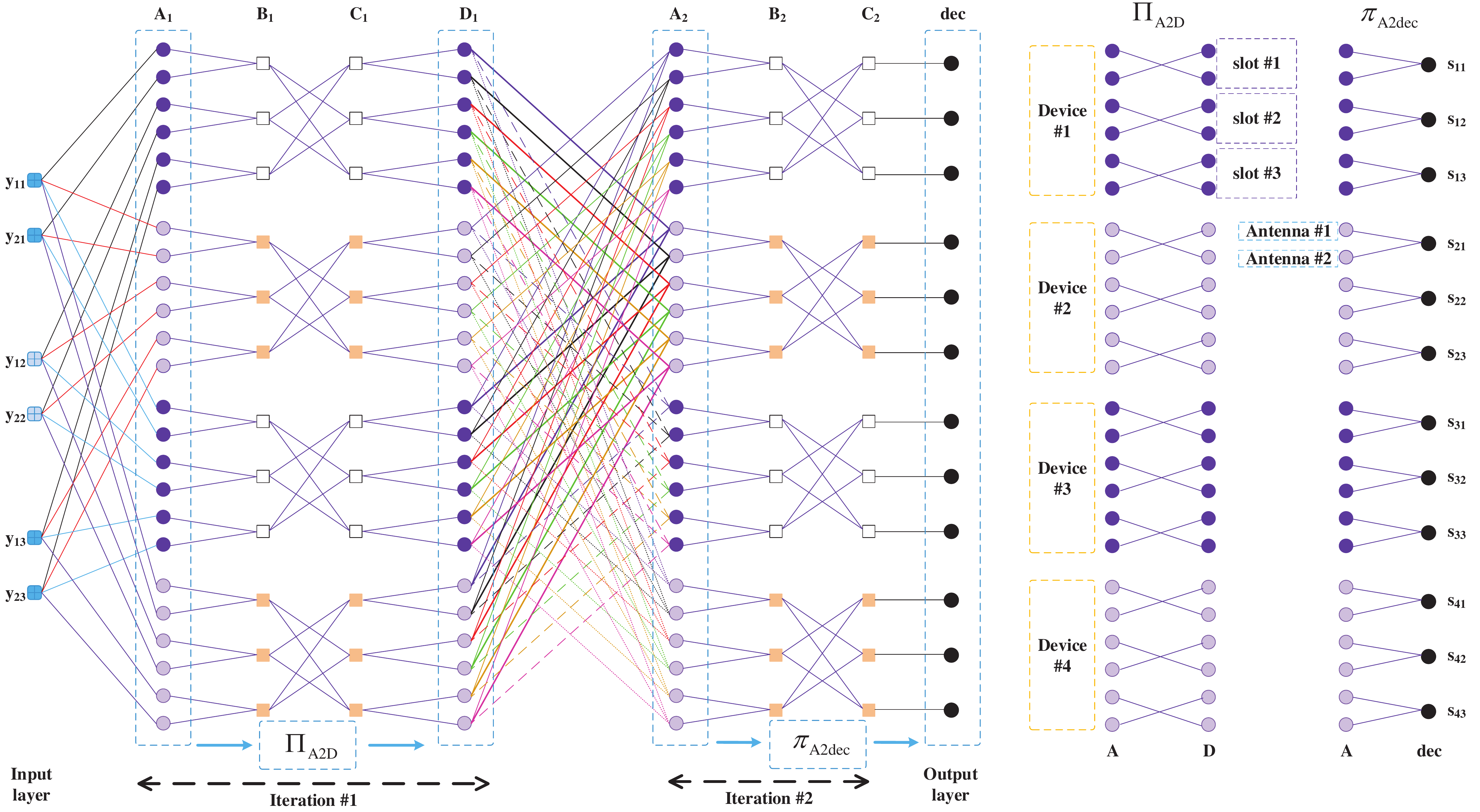}\\
	\caption{Neural network structure for the DNN-MP-AD algorithm ($M=2$, $N_s=4$, $N_p=3$ and $L=2$ iterations).}\label{DNN}
\end{figure*}

For a better understanding of the DNN-MP-AD algorithm, an example is illustrated in Fig. \ref{DNN} assuming that related parameters are $M=2$, $N_s=4$, $N_p=3$ and $L=2$ iterations for the MP-AD algorithm. Entries in the received signal matrix $\mathbf{Y}$ and the channel matrix $\mathbf{H}$ now serve as nodes of the input layer. However, entries in $\mathbf{H}$ are not depicted in Fig. \ref{DNN} for the clarity of presentation. Each complete iteration of the MP-AD algorithm is now transformed into four different hidden layers, corresponding to the four message update processes in Fig. \ref{updatediagram}. The output layer, i.e., the decision layer has $N_s \times N_p=12$ nodes, corresponding to the LLR messages for the $12$ entries in $\mathbf{S}$.
To be specific, the messages passing from SNs to VNs in Fig. \ref{updatediagram}(a) are represented by the nodes in hidden layer A, which indicates that there are $M\times N_s\times N_p =24$ nodes in hidden layer A. The messages passing from VNs to CNs in Fig. \ref{updatediagram}(c) and the messages passing from CNs to VNs in Fig. \ref{updatediagram}(b) are represented by the nodes in hidden layers B and C, respectively. Therefore, the number of nodes in hidden layers B and C is $N_s \times N_p = 12$. The messages passing from VNs to SNs in Fig. \ref{updatediagram}(d) are represented by the nodes in hidden layer D. Consequently, there are $M\times N_s\times N_p =24$ nodes in hidden layer D. We further employ the subscript as the index for different iterations. Then the nodes in hidden layer $\text{D}_{i}$ serve as the input for the next hidden layer $\text{A}_{i+1}$. It should be noted that, according to (\ref{SUMNODE}), the input for subsequent hidden layers A also includes the nodes in the input layer. However, for the clarity of presentation in Fig. \ref{DNN}, we do not depict the connection from the input layer to the hidden layers $\text{A}_{i}$ with $i>1$ since this connection is identical to that from the input layer to $\text{A}_{1}$. Furthermore, the output layer does not require the nodes from hidden layer D, which indicates that the last iteration only contains 3 hidden layers in Fig. \ref{DNN}. Finally, according to (\ref{SDvnupdate}) and (\ref{var_dec_LLR}), the nodes in hidden layer A are required by hidden layer D and the output layer. The connections from hidden layer A to hidden layer D and the output layer are illustrated by two interleavers ${\Pi _{{\rm{A2D}}}}$ and ${\pi_{A2dec}}$ in Fig. \ref{DNN}.

For the convenience of illustration, the nodes are colored to indicate the node affiliation in Fig. \ref{DNN}. To be specific, nodes in different layers are divided into $N_s=4$ groups according to the device they belong to. Within each device group, the nodes in hidden layers A and D are divided into $N_p=3$ subgroups according to different slots, while the $M=2$ nodes within each slot subgroup are distinguished by the antenna they correspond to. In terms of the hidden layers B, C and the output layer, the nodes in each device group are simply distinguished according to the slot. With this node affiliation, the $m$-th node in the $p$-th slot subgroup of the $s$-th device group in hidden layer $\text{A}_{\tau}$ is interpreted as the LLR message $l_{mp \to sp}^s(\tau)$ in Fig. \ref{updatediagram}(a). Similarly, we can define the interpretation for the nodes in other hidden layers and the connections between different layers are uniquely determined as in Fig. \ref{DNN}. It is noted that, different from the typical DNN, the neural network structure for the DNN-MP-AD algorithm exhibits relatively sparse inter-layer connection.

Overall, there are $4L-1$ hidden layers and one input layer as well as one output layer in the neural network structure for the DNN-MP-AD algorithm, where $L$ is the number of iterations for the iterative MP-AD algorithm.
\subsection{Weighted Message Update for DNN-MP-AD Algorithm}
Based on the neural network structure, the weighted message update at different hidden layers of the DNN-MP-AD algorithm is presented as follows. For notational convenience, we denote $\mathcal{W}$ as the set of all the weights in the following equations.
\subsubsection{Hidden Layer A}
As stated, the nodes in hidden layer A represent the messages passing from SNs to VNs. We further make some modifications on (\ref{SUMNODE}) for the convenience of presentation and present the weighted message update equation at hidden layer A as follows.
\begin{equation}\label{hiddenlayerA}
l_{spm}^{{{\rm{A}}_i}} = \dfrac{{2(w_{spm}^y{y_{mp}} - u_{spm}^{*w}){h_{ms}} - h_{ms}^2}}{{2v_{spm}^{*w}}}
\end{equation}
with
\begin{equation}\label{AUV}
\begin{split}
&u_{spm}^{*w}\!\! = \!\!\!\!\sum\limits_{t \in {N_{{\rm{D2A}}}}(spm)}\!\!\!\!\!\!\!\!\!\!{w_{{s^*}spm}^u{h_{m{s^*}}}{{(1 \!+\! {e^{ - l_{{s^*}pm}^{{{\rm{D}}_{i - 1}}}}})}^{ - 1}}} \\
&v_{spm}^{*w} \!\!=\!\!\!\!\sum\limits_{t \in {N_{{\rm{D2A}}}}(spm)}\!\!\!\!\!\!\!\!\!\!{w_{{s^*}spm}^vh_{m{s^*}}^2\!{{(2 \!+\! {e^{ - l_{{s^*}pm}^{{{\rm{D}}_{i - 1}}}}} \!+\! {e^{l_{{s^*}pm}^{{{\rm{D}}_{i - 1}}}}})}^{\!\! -\! 1}}}\!\!\!\!\!\!+\! w_{spm}^{\sigma^2}\sigma_n^2
\end{split}
\end{equation}
where $i$ is the iteration index and the node $spm$ in hidden layer A corresponds to the $m$-th antenna in the $p$-th slot subgroup of the $s$-th device group. $N_{\text{D2A}}(spm)$ is the set of neighbors in the previous layer $D_{i-1}$, i.e., nodes in $N_{\text{D2A}}(spm)$ are connected to node $spm$ in hidden layer A while $s^*$ is the index of the device group for neighbor $t$ in hidden layer $D_{i-1}$. The weights $w_{spm}^y, w_{{s^*}spm}^u, w_{{s^*}spm}^v$ and $w_{spm}^{\sigma^2}$ are imposed on the messages ${y_{mp}}, u_{spm}^{*w}, v_{spm}^{*w}$ and $\sigma_n^2$ respectively while $\sigma_n^2$ is regarded as the bias term of hidden layer A. For notational simplicity, the iteration index is removed from the weights but these weights are independently trained across different iterations. It should be noted that at the first iteration, $l_{s^*pm}^{\text{D}_{0}}$ is assumed to be 0 for all the nodes in hidden layer $\text{D}_{0}$ since no priori information is passed from VNs to SNs at the beginning.
\subsubsection{Hidden Layer B}
The nodes in hidden layer B represent the messages passing from VNs to CNs in (\ref{SDvnupdate}). The weighted update equation is now rewritten as
\begin{equation}\label{hiddenlayerB}
l_{sp}^{{{\rm{B}}_i}} = \sum\limits_{t \in {N_{{\rm{A2B}}}}(sp)} {w_{sp{m^*}}^{{\rm{A2B}}}l_{sp{m^*}}^{{{\rm{A}}_i}}}  + \overline{w}_{sp}^{B}{\overline l _{sp}}
\end{equation}
where $w_{sp{m^*}}^{{\rm{A2B}}}$ and $\overline{w}_{sp}^{B}$ are weights while ${\overline l _{sp}}$ is regarded as the bias term of hidden layer B. Node $sp$ in hidden layer B corresponds to the $p$-th slot of the $s$-th device group and $N_{\text{A2B}}(sp)$ is the set of neighbor nodes in hidden layer A for node $sp$ in hidden layer B. In addition, $m^*$ is the index of the antenna which the neighbor $t$ in  $N_{\text{A2B}}(sp)$ corresponds to.
\subsubsection{Hidden Layer C}
The nodes in hidden layer C represent the messages passing from CNs to VNs in (\ref{check_update_LLR}). Therefore, the update equation is now rewritten as
\begin{equation}\label{hiddenlayerC}
l_{sp}^{{{\rm{C}}_i}} =  - \log ({e^{ - \widetilde l_{sp}^{{{\rm{C}}_i}}}} - 1)
\end{equation}
with
\begin{equation}\label{subC}
\widetilde l_{sp}^{{{\rm{C}}_i}} = w_{sp}^{p_a}\log ({p_a}) - \sum\limits_{t \in {N_{{\rm{B2C}}}}(sp)} {w_{s{p^*}p}^{{\rm{B2C}}}\log ({e^{l_{sp}^{{{\rm{B}}_i}}}} + 1)}
\end{equation}
where $w_{sp}^{p_a}$ and $w_{s{p^*}p}^{{\rm{B2C}}}$ are weights and $\log ({p_a})$ is regarded as the bias term of hidden layer C. Node $sp$ in hidden layer C corresponds to the $p$-th slot of the $s$-th device group and $N_{\text{B2C}}(sp)$ is the set of neighbor nodes in hidden layer B for node $sp$ in hidden layer C. In addition, $p^*$ is the index of the slot which the neighbor $t$ in  $N_{\text{B2C}}(sp)$ corresponds to.
\subsubsection{Hidden Layer D}
As shown in Fig. \ref{DNN}, the input of hidden layer D includes the nodes in both hidden layer A and hidden layer C. Therefore, we have
\begin{equation}\label{hiddenlayerD}
l_{spm}^{{{\rm{D}}_i}} = \sum\limits_{t \in {N_{{\rm{A2D}}}}(spm)} {w_{sp{m^*}m}^{{\rm{A2D}}}l_{sp{m^*}}^{{{\rm{A}}_i}}}  + w_{spm}^{{\rm{C2D}}}l_{sp}^{{{\rm{C}}_i}} + \overline{w}_{sp}^{D}{\overline l _{sp}}
\end{equation}
where $w_{sp{m^*}m}^{{\rm{A2D}}}, w_{spm}^{{\rm{C2D}}}$ and $\overline{w}_{sp}^{D}$ are weights while ${\overline l _{sp}}$ is regarded as the bias term of hidden layer D. Node $spm$ in hidden layer D corresponds to the $m$-th antenna in the $p$-th slot subgroup of the $s$-th device group. $N_{\text{A2D}}(spm)$ is the set of neighbor nodes in hidden layer A for node $spm$ in hidden layer D and $m^*$ is the index of the antenna which the neighbor $t$ in $N_{\text{A2D}}(spm)$ corresponds to.
\subsubsection{Output Layer}
Similar to the hidden layer D, the input of the output layer includes the nodes in the last hidden layer $\text{A}_L$ and the last hidden layer $\text{C}_L$.
\begin{equation}\label{outputlayer}
l_{sp}^{dec} = \sum\limits_{t \in {N_{{\rm{A2dec}}}}(sp)} {w_{sp{m^*}}^{{\rm{A2dec}}}l_{sp{m^*}}^{{{\rm{A}}_{L}}}}  + w_{sp}^{{\rm{C2dec}}}l_{sp}^{{{\rm{C}}_{L}}} + \overline{w}_{sp}^{dec}{\overline l _{sp}}
\end{equation}
where $w_{sp{m^*}}^{{\rm{A2dec}}}, w_{sp}^{{\rm{C2dec}}}$ and $\overline{w}_{sp}^{dec}$ are weights while ${\overline l _{sp}}$ is regarded as the bias term of the output layer. Node $sp$ in the output layer is the LLR for the $sp$-th entry of the matrix $\mathbf{S}$, $N_{\text{A2dec}}(sp)$ is the set of neighbor nodes in the last hidden layer A for node $sp$ in the output layer and $m^*$ is the index of the antenna which neighbor $t$ in $N_{\text{A2dec}}(sp)$ corresponds to. Finally, the decision is made as in (\ref{zerone}).
\subsection{Loss Function}
In order to accelerate the training of the weights in the DNN, we utilize the cross entropy function as the loss function. It is noted that the relationship between the output LLR $l_{sp}^{\text{dec}}$ and the non-zero probability ${p_{sp}^{dec}}$ of the corresponding VN $s_{sp}$ coincides with the sigmoid function, i.e., ${p_{sp}^{dec}} = {(1 + {e^{ - l_{sp}^{\text{dec}}}})^{ - 1}}$. According to this relationship, the non-zero probability vector $\textbf{p}^{dec} = [{p_{11}^{dec}}, \ldots ,{p_{{N_s}1}^{dec}};{p_{12}^{dec}}, \ldots ,{p_{{N_s}2}^{dec}}; \cdots ;{p_{1{N_p}}^{dec}}, \ldots ,{p_{{N_s}{N_p}}^{dec}}]$ and the true value vector, i.e., the \emph{label vector} $\textbf{s} = [{s_{11}}, \ldots ,{s_{{N_s}1}};{s_{12}}, \ldots ,{s_{{N_s}2}}; \cdots ;{s_{1{N_p}}}, \ldots ,{s_{{N_s}{N_p}}}]$ are taken to calculate the cross entropy
\begin{equation}\label{lossfunction}
\text{loss} = \frac{1}{N_sN_p}\sum_{\{s,p\}}\left[ { s_{sp} \ln p_{sp}^{dec} + (1 - s_{sp}) \ln (1 - p_{sp}^{dec})} \right]
\end{equation}

In order to further accelerate the training of the weights near the input layer, the loss function is modified by introducing the multi-loss, i.e., we consider the non-zero probability of each middle iteration $\textbf{p}^{i} = [{p_{11}^{i}}, \ldots ,{p_{{N_s}1}^{i}};{p_{12}^{i}}, \ldots ,{p_{{N_s}2}^{i}}; \cdots ;{p_{1{N_p}}^{i}}, \ldots ,{p_{{N_s}{N_p}}^{i}}]$.
\begin{equation}\label{fulllossfunction}
\text{loss}_\text{full} \!=\! \frac{1}{N_sN_p}\sum_i\sum_{\{s,p\}}\left[ { s_{sp} \ln p_{sp}^{i} + (1 \!-\! s_{sp}) \ln (1 \!-\! p_{sp}^{i})} \right]
\end{equation}
where the iteration index $i \in \{1, 2, \ldots, L-1, dec\}$ and the non-zero probability $p_{sp}^{i}$ is related to the output LLR of each middle iteration, i.e., ${p_{sp}^{i}} = {(1 + {e^{ - l_{sp}^{i}}})^{-1}}$.

Finally, the weights in the set $\mathcal{W}$ are trained to minimize the loss function in (\ref{fulllossfunction}). After the training process, these weights are preserved in the neural network and employed by the DNN-MP-AD algorithm for future inputs. The procedure of the DNN-MP-AD algorithm is almost identical to that of the MP-AD algorithm in Phase 1 of Scheme \ref{alg}. The differences are that $\mathcal{W}$ is also included as the input information and equations (\ref{hiddenlayerA}), (\ref{hiddenlayerB}), (\ref{hiddenlayerC}) and (\ref{hiddenlayerD}) are employed for the four message updates in each iteration while (\ref{outputlayer}) is employed for the output message.
\section{Simulation Results}\label{sec:simulation}
\begin{table}
	\scriptsize
	\renewcommand\arraystretch{1.5}
	\caption{Simulation configurations}
	\centering
	\begin{tabular}{|c|c|}
		\hline
		Simulation platform for MP-AD algorithm& Matlab 2018a\\
		Simulation platform for DNN-MP-AD algorithm& Tensorflow 1.6 Python 3.5\\
		GPU& GTX 1080 Ti\\
		Optimizer& AdamOptimizer\\
		Training set size & $2\times 10^6$\\
		Test set size & $5\times 10^5$\\
		Mini-batch size   & $2 \times 10^3$\\
		Epoch number & 20\\
		Initial weight & 1\\
		Learning rate & $1\times 10^{-3}$\\
		\hline
	\end{tabular}
	\label{simuconfig}
\end{table}
\begin{figure}
	\centering
	\includegraphics[width=0.45\textwidth]{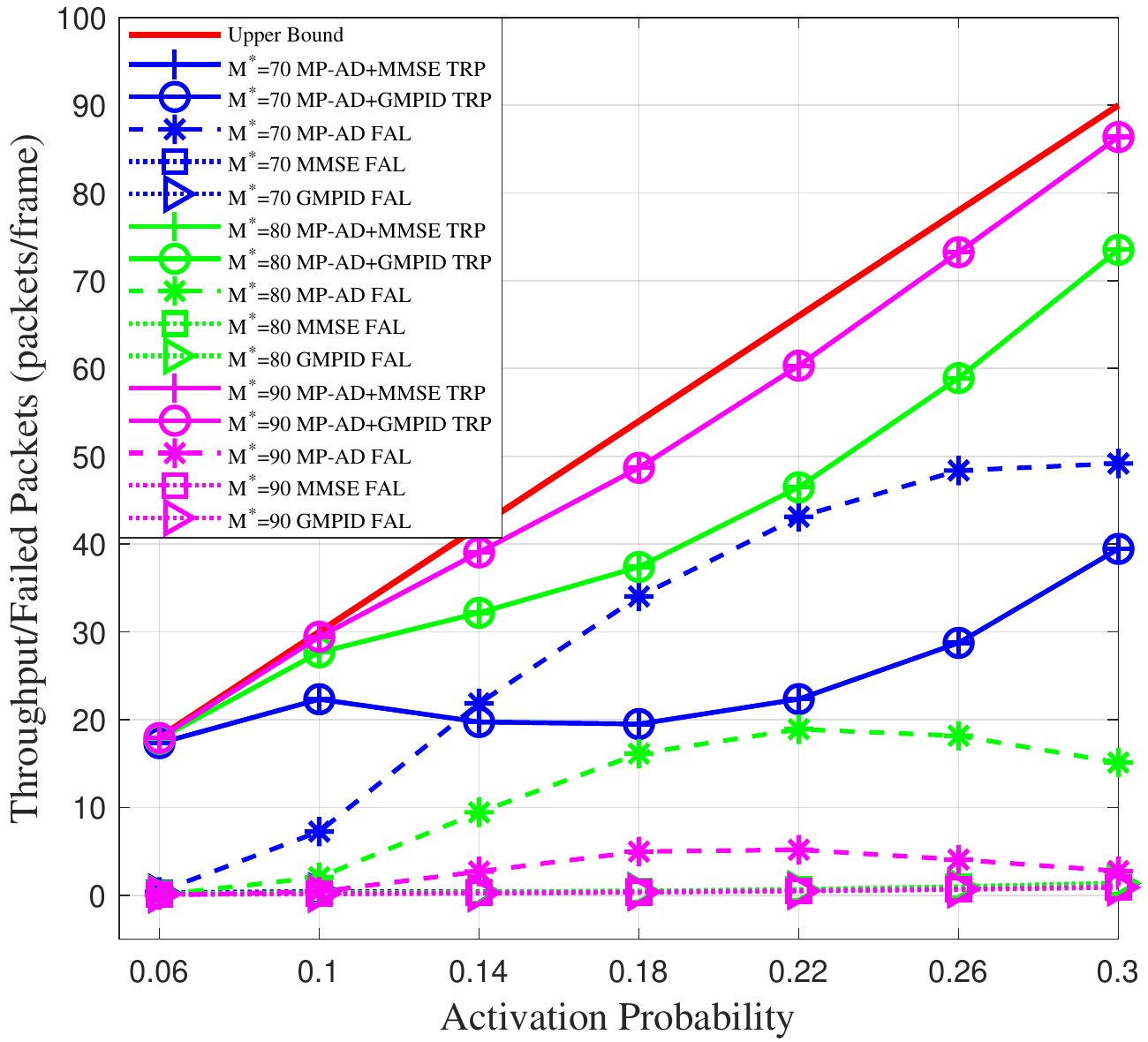}\\
	\caption{Throughput (denoted by TRP) and failed packets (denoted by FAL) of the proposed fixed-symbol aided RA scheme with various activation probability $p_a$ and $M^*$. Other related parameters are $N_s=300$, $N_p=6$, $\epsilon_{\text{rec}}=2\times10^{-4}$ and SNR = 25dB. The MP-AD algorithm is performed with 10 iterations while the GMPID for MUD is performed with 30 iterations in the proposed fixed-symbol aided RA scheme. The simulation results are averaged over $10^5$ repeated blocks.}\label{throughput}\vspace{-0.3cm}
\end{figure}
For the simulations, we consider the complex-valued Rayleigh fading channel. As stated in Remark \ref{realcomplex}, the complex-valued scenario is actually equivalent to the real-valued scenario for the MP-AD algorithm and the DNN-MP-AD algorithm while the only difference is that $M$ antennas in the real-valued scenario corresponds to $M^*=M/2$ antennas in the complex-valued scenario. Configurations related to the following simulations are listed in Table \ref{simuconfig}.
\begin{figure*}
	\centering
	\includegraphics[width=1\textwidth]{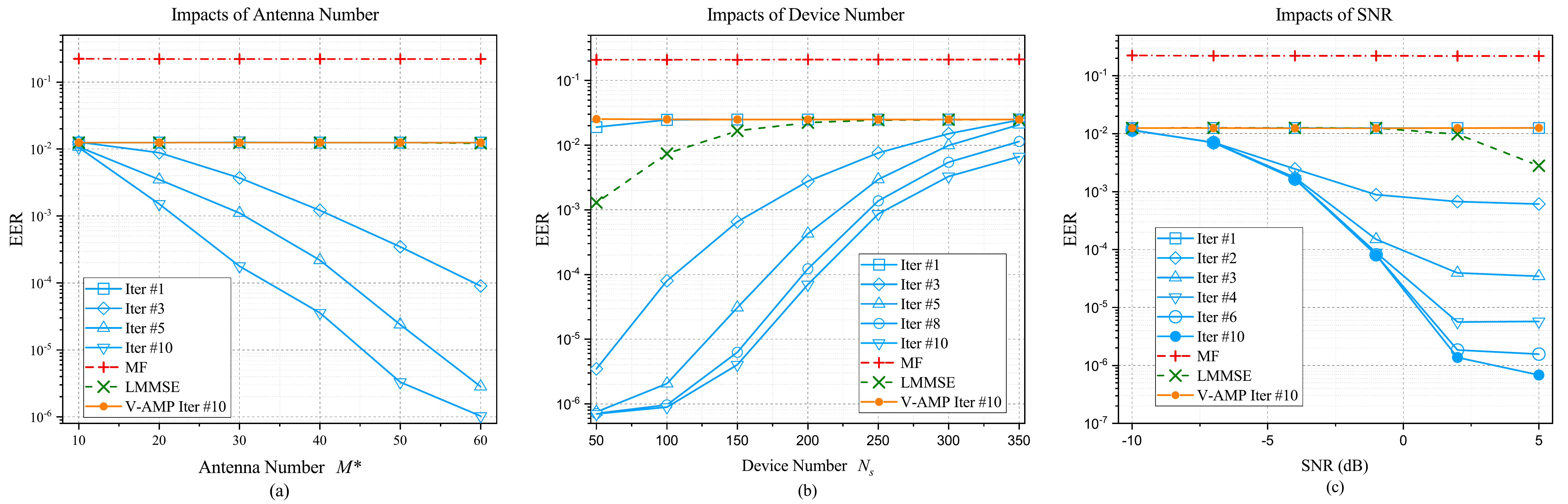}
	\caption{Simulation results for the EER performance of the MP-AD algorithm under different settings (a) various $M^*$ and $N_s=200$, $N_p=4$, SNR = 0 dB, $p_a=0.05$ (b) various $N_s$ and $M^*=50$, $N_p=4$, SNR = 0 dB and $p_a=0.1$ (c) various SNR and $M^*=35$, $N_s=70$, $N_p=4$, $p_a=0.05$. The entries in the channel matrix $\mathbf{H}$ are assumed \emph{i.i.d} with Rayleigh distribution $\mathcal{CN}(0,1)$.}\label{simulation}\vspace{-0.3cm}
\end{figure*}
	\subsection{Throughput of Fixed-Symbol Aided RA Scheme}
The throughput is defined as the average number of successfully received data packets in each RA frame. As stated above, the data symbols are assumed Gaussian distributed with zero mean and variance $P_{\text{r}}$, where the mean received power $P_{\text{r}}$ after reverse power control is set to 1. The signal-to-noise ratio (SNR) is defined as $\text{SNR}=10\log_{10}{\dfrac{P_{\text{r}}}{\sigma_n^2}}$ where $\sigma_n^2$ is the variance of the complex additive white Gaussian noise. For the proposed fixed-symbol aided RA scheme, the MP-AD algorithm is employed to detect the device activity in each slot while the MMSE estimator or the GMPID \cite{Lei2016Convergence} are employed to recover the Gaussian data symbols. The Gaussian data symbols are considered correctly recovered as long as the recovery Mean Square Error (MSE) is lower than a predefined threshold $\epsilon_{\text{rec}}$. The value of $\epsilon_{\text{rec}}$ is determined by the realistic coding scheme for the Gaussian symbol. For the sake of simplicity, we assume that the threshold $\epsilon_{\text{rec}}$ is determined as the lowest value such that the MMSE estimator could correctly recover the Gaussian data symbol for the given SNR. Then, a data packet from an activated device is assumed successfully received by the BS if and only if the activity of all the devices is correctly detected in the corresponding slot and the Gaussian data symbols are correctly recovered.

The throughput of the proposed fixed-symbol aided RA scheme is shown in Fig. \ref{throughput}, where the throughput upper bound is obtained by assuming that the data packets from all the activated devices are successfully received, i.e., the upper bound is equal to $N_s\times p_a$. In addition, we also illustrate the number of failed packets caused by the error of the MP-AD algorithm, the MMSE estimator and the GMPID. It is shown by the solid curves that for different activation probability $p_a$, an increasing number of antennas could greatly improve the throughput of the proposed RA scheme. In addition, the throughput of the RA scheme with GMPID employed for MUD is identical to that with the MMSE estimator employed for MUD. The dashed curves show that, for different number of antennas, the failed packets in the fixed-symbol aided RA scheme are mainly caused by the error of the MP-AD algorithm, indicating that the device activity detection is the key to the throughput of the proposed RA scheme. Finally, the dotted curves show that the low-complexity GMPID could well approach the performance of the MMSE estimator and excellent MUD performance can be guaranteed as long as the device activity detection is accurate.

\subsection{Impacts of System Parameters on Detection Accuracy}
According to the simulation results in Fig. \ref{throughput}, the accuracy of the MP-AD algorithm is the key to the throughput of the proposed fixed-symbol aided RA scheme. Therefore, we focus on the accuracy of the MP-AD algorithm for the following simulations and investigate the impacts of different system parameters. It is noted that when false alarm (FA) of the MP-AD algorithm occurs, i.e., an inactive device is considered active, the recovered data symbol of this FA device is close to zero according to the GMPID. However, since Gaussian data symbols are considered, the recovered symbol of this FA device is also accepted by the BS, leading to the wrong reception of inactive devices. Therefore, both missed detection (MD) and FA are not acceptable in the fixed-symbol aided RA scheme and we do not further distinguish these two types of errors. Instead, we employ the element error rate (EER) for every element of matrix $\mathbf{S}$ to indicate the detection accuracy.

The simulation results for the impacts of three different parameters $M^*$, $N_s$ and SNR are illustrated in Fig. \ref{simulation}, respectively. For comparison, the performances of the Linear Minimum Mean Square Error (LMMSE) estimator \cite{LMMSE}, the matched filter (MF) estimator and the V-AMP estimator \cite{AMPmassive} are also investigated for the problem in (\ref{modelmatrix}) while additional check constraint is imposed on their final hard decision, i.e., at most one element in each row of matrix $\mathbf{S}$ is 1.

The impact of antenna number $M^*$ is illustrated in Fig. \ref{simulation}(a). It can be observed that the EER performance improves almost linearly in the logarithmic scale with increasing $M^*$. Therefore, an accurate detection can be guaranteed within a feasible number of iterations as long as the $N_s$ devices are supported by enough antennas. This observation can be explained by the fact that more antennas bring more incoming messages for each VN according to (\ref{var_dec_LLR}). As a result, the final decision can be made with higher accuracy.

Similar observations can be found in Fig. \ref{simulation}(b) for the impact of device number $N_s$. When a small number of devices are supported by the massive MIMO BS, e.g. $N_s=50$, the MP-AD algorithm exhibits higher accuracy with fewer iterations. However, the detection accuracy deteriorates as the number of devices increases.
\begin{figure*}
	\centering
	\includegraphics[width=1\textwidth]{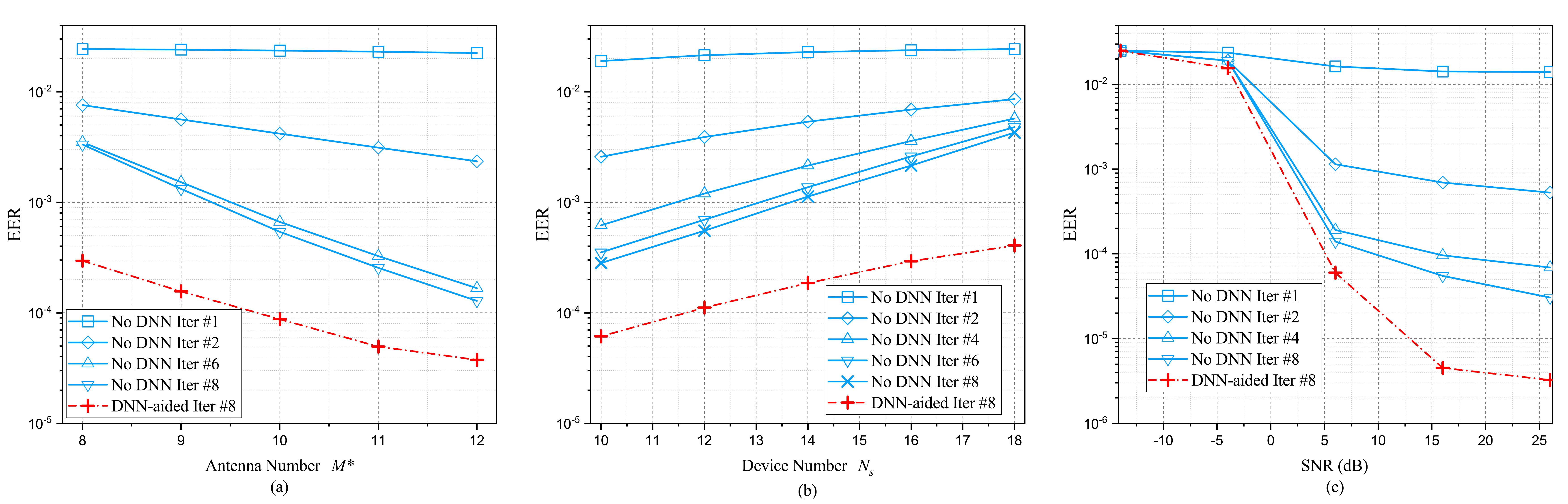}
	\caption{Simulation results for the improvement of the DNN-MP-AD algorithm under different settings (a) various $M^*$ and $N_s=20$, $N_p=4$, SNR = 16 dB, $p_a=0.1$ (b) various $N_s$ and $M^*=7$, $N_p=4$, SNR = 16 dB and $p_a=0.1$ (c) various SNR and $M^*=10$, $N_s=10$, $N_p=4$, $p_a=0.1$. The entries in the channel matrix $\mathbf{H}$ are assumed \emph{i.i.d} with Rayleigh distribution ${\cal CN}(0,1)$.}\label{DNNsimulation}\vspace{-0.3cm}
\end{figure*}
The impact of SNR is illustrated in Fig. \ref{simulation}(c). It is shown that as SNR increases, the EER performance of the MP-AD algorithm gets improved.

As shown in Fig. \ref{simulation}, the MP-AD algorithm outperforms the other three estimators in terms of the detection accuracy. Specifically, the matrix $\mathbf{S}$ follows a different distribution from the estimation target in \cite{AMPmassive}. Therefore, the V-AMP estimator for sparse signal recovery does not work since it fails to consider the transmission constraint in each row of $\mathbf{S}$. In addition, according to the simulations, at most 10 iterations are required by the MP-AD algorithm to reach the fixed EER point, which proves the implementation feasibility of the MP-AD algorithm.
\subsection{Improvement by DNN-MP-AD Algorithm}
For training convenience, we focus on small-scale system models. In addition, only 8 iterations are considered for the DNN-MP-AD algorithm since it is verified in previous simulations that less than 10 iterations are required by the MP-AD algorithm to reach the fixed EER point.

Similar to the simulations in Fig. \ref{simulation}, we also investigate the performance improvement brought by the DNN-MP-AD algorithm from three different perspectives. The results are illustrated in Fig. \ref{DNNsimulation}. It is shown in Fig. \ref{DNNsimulation} that under most settings considered, the DNN-MP-AD algorithm could reduce the EER of the MP-AD algorithm by almost one order of quantity. That is, the DNN structure brings significant improvement to the detection accuracy without causing any additional online computational complexity. According to Fig. \ref{DNNsimulation}(a) and Fig. \ref{DNNsimulation}(b), the performance improvement of the DNN-MP-AD algorithm is more prominent for overloaded systems, i.e., when $M^*$ is small or $N_s$ is large. In addition, it is shown in Fig. \ref{DNNsimulation}(c) that the DNN-MP-AD algorithm could further lower the EER in the high SNR regime. These observations are consistent with the assumption of the correlation problem caused by the CN update in Section \ref{correlation}.
\subsection{Robustness Against Channel Variation}\label{variation}
As stated above, for stationary devices with fixed locations, the CSI remains static within a certain period and exhibits a strong time correlation for a longer time. The channel variation due to the diminishing time correlation can be modelled as the channel estimation error for the MP-AD algorithm and the DNN-MP-AD algorithm. In addition, the estimated CSI matrix may be inevitably contaminated by estimation errors due to the massiveness of M2M communications. Therefore, we investigate the robustness of the MP-AD algorithm and the DNN-MP-AD algorithm against channel estimation errors, i.e., channel variation via the following simulations.

Some specific assumptions are stated as follows. The channel estimation error matrix $\mathbf{H_e}$ is assumed independent from $\mathbf{H}$. The entries in $\mathbf{H_e}$ are \emph{i.i.d} and Rayleigh distributed with zero mean and variance $\sigma_e^2$ while the entries in the channel matrix $\mathbf{H}$ are assumed \emph{i.i.d} and Rayleigh distributed with zero mean and unit variance. As a result, the CSI matrix acquired at the BS is $\mathbf{H+H_e}$, which is further employed for the MP-AD algorithm and the DNN-MP-AD algorithm.
\subsubsection{Robustness of the MP-AD algorithm}
\begin{figure}
	\centering
	\includegraphics[width=0.44\textwidth]{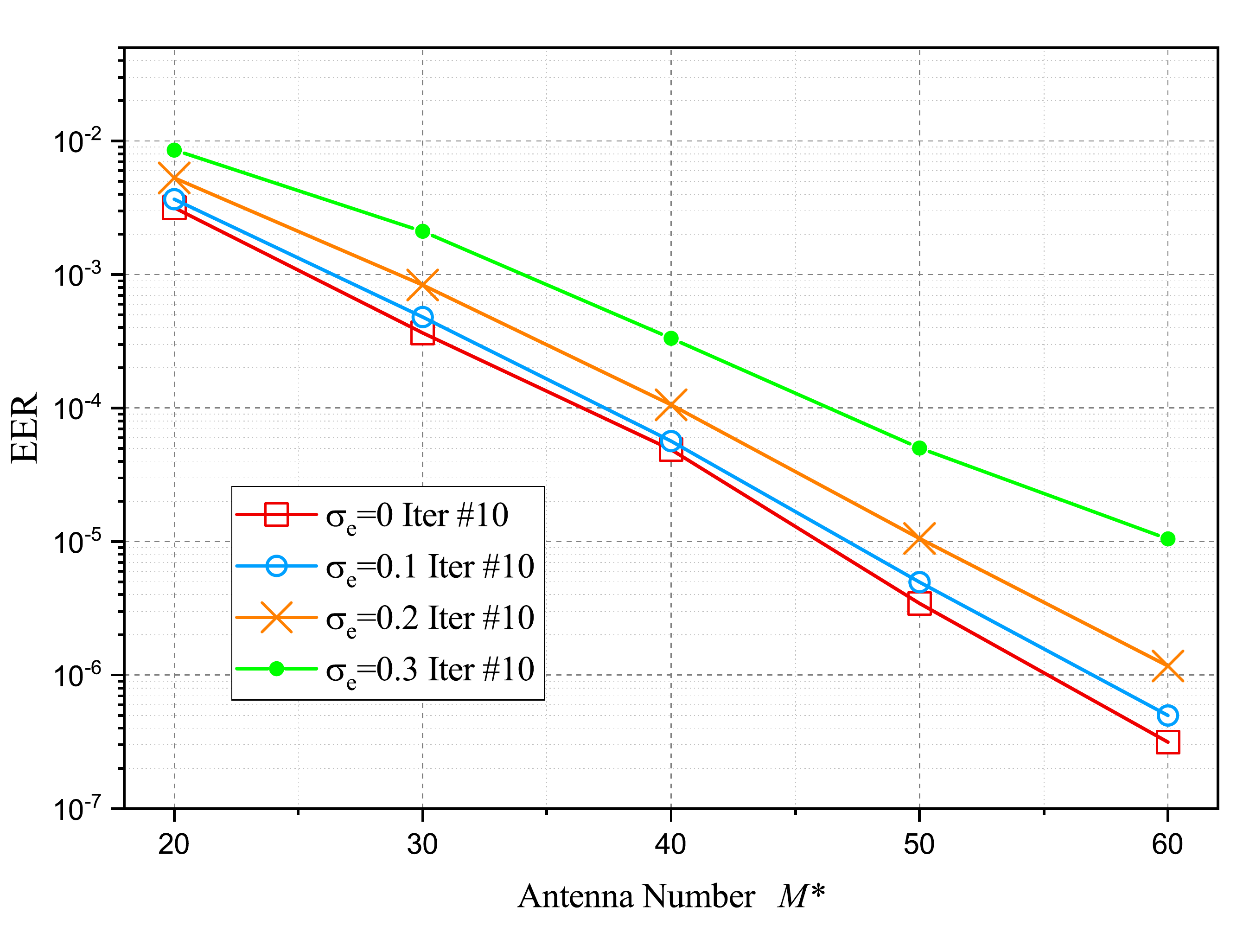}
	\caption{Simulations for the robustness of the MP-AD algorithm with various $M^*$ and $N_s=150$, $N_p=4$, $p_a=0.1$ and SNR = 0 dB.}\label{HeLargeM}\vspace{-0.3cm}
\end{figure}

The EER performances of the MP-AD algorithm with various number of antennas and different $\sigma_e$'s are illustrated in Fig. \ref{HeLargeM}. It is shown that when the variance of the estimation error is relatively small, e.g., $\sigma_e=0.1$ or $0.2$, the performance degradation is almost negligible. However, the detection accuracy is undermined seriously with a larger channel estimation error, e.g., $\sigma_e=0.3$. In addition, according to Fig. \ref{HeLargeM}, more antennas can improve the EER performance but also lead to more serious performance degradation when $\sigma_e$ is fixed.
\subsubsection{Robustness of the DNN-MP-AD algorithm}
\begin{figure}
	\centering
	\includegraphics[width=0.44\textwidth]{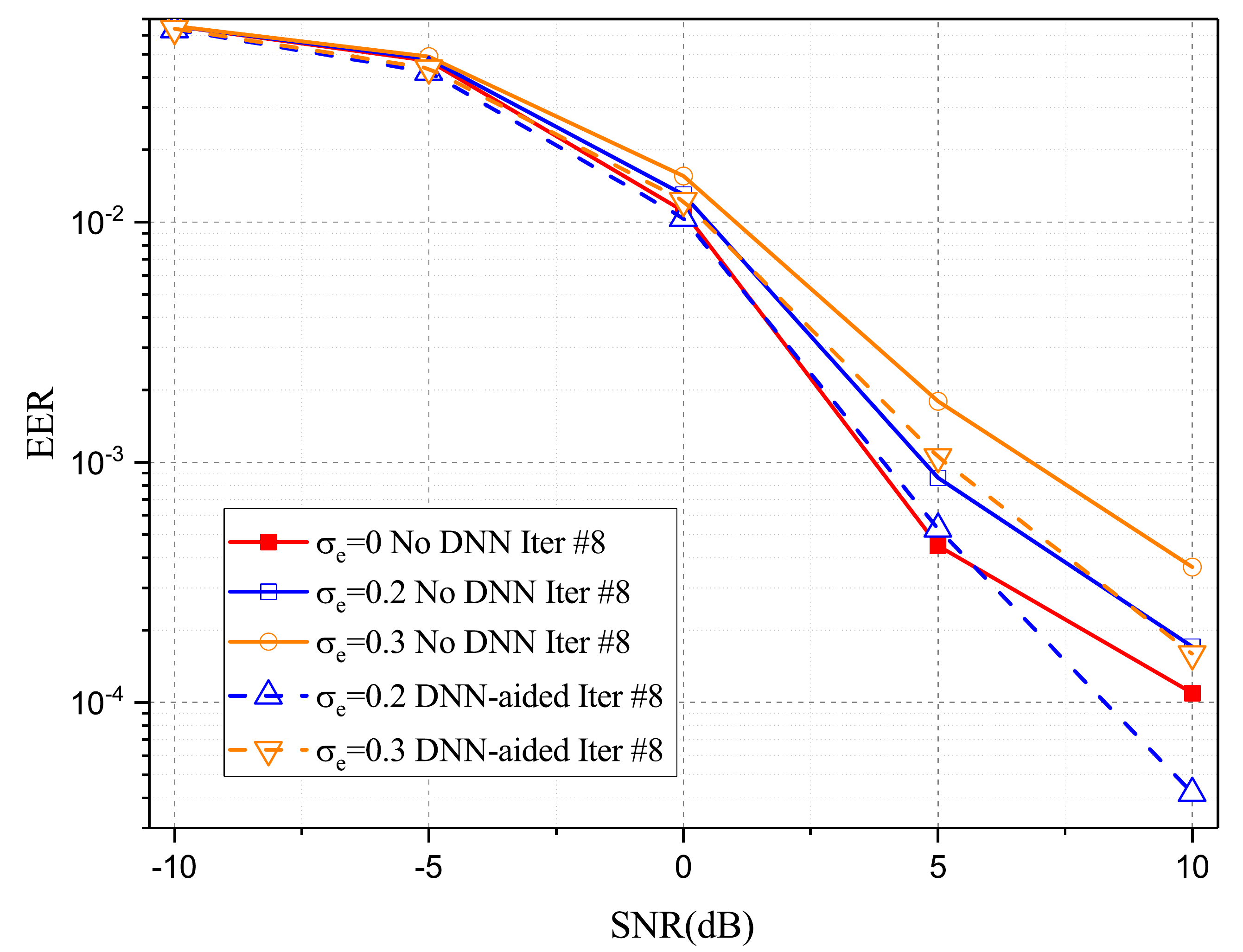}
	\caption{Simulations for the robustness of the DNN-MP-AD algorithm with various SNR and $M^*=10$, $N_s=10$, $N_p=3$, and $p_a=0.2$.}\label{HeDNN}\vspace{-0.3cm}
\end{figure}
Again, we consider small-scale system models and 8 iterations of the DNN-MP-AD algorithm for the convenience of training. The simulation results are illustrated in Fig. \ref{HeDNN} while the performances of the MP-AD algorithm are also included for comparison. It is shown that the training gain of the DNN could compensate for the impacts of channel estimation errors. Specifically, when $\sigma_e=0.2$, the EER performance of the DNN-MP-AD algorithm is equal or even superior to that of the MP-AD algorithm with perfect CSI. Similarly, the EER performance is also prominently improved with the DNN structure when $\sigma_e=0.3$.
\section{Conclusion and Future Work}\label{sec:conclusion}
A fixed-symbol aided RA scheme was proposed for crowded M2M communications. In order to deal with the issue of severe collisions, one fixed symbol is inserted into each transmitted data packet and we designed the MP-AD algorithm for the device activity detection, so that MUD could be further employed to decode collided data packets in each slot. The DNN-MP-AD algorithm was further designed to alleviate the correlation problem. Finally simulation results verified the throughput of the proposed RA scheme, the accuracy of the MP-AD algorithm as well as the improvement by the DNN structure in crowded M2M communications.

Future work may be focused on the analysis of the proposed MP-AD algorithm to obtain the bound of the detection probability. The influences of  different network parameters, such as the activation probability, the number of antennas and devices, and SNR, on the convergence of the iterative MP-AD algorithm should be investigated theoretically. Aided by this analysis, the throughput bound of the proposed fixed-symbol aided RA scheme can be further provided.
\bibliographystyle{IEEEtran}


\end{document}